\documentclass[preprint,prd,aps,floatfix,preprintnumbers,superscriptaddress,bibnotes,nofootinbib]{revtex4-1}

\usepackage{epsfig}
\usepackage{amsmath}
\usepackage{latexsym}
\usepackage[psamsfonts]{amssymb}
\usepackage{graphicx}
\usepackage{ulem}
\usepackage{longtable}
\usepackage{epstopdf}
\usepackage{bm}
\usepackage{color}

\newcommand{\be}{\begin{equation}}
\newcommand{\ee}{\end{equation}}
\newcommand{\bea}{\begin{eqnarray}}
\newcommand{\eea}{\end{eqnarray}}
\newcommand{\bi}{\begin{itemize}}
\newcommand{\ei}{\end{itemize}}

\begin{document}

\title{
Glueball mass spectrum at finite temperature 
revisited: \\
Constant contribution in glueball correlators \\
in the deconfinement phase
}


\author{Toshizo Arikawa}\email{toshizo.arikawa.s3@alumni.tohoku.ac.jp}
\author{Keita Sakai}\email{keita.sakai.s4@alumni.tohoku.ac.jp}
\author{Shoichi Sasaki}\email{ssasaki@nucl.phys.tohoku.ac.jp}

\affiliation{Department of Physics, Tohoku University, Sendai 980-8578, Japan}

\date{\today}
\begin{abstract}
We study the glueball properties at finite temperature from 
the temporal correlation in $SU(3)$ Yang-Mills theory using anisotropic lattice QCD. 
Although the existence of a constant contribution to the meson correlation function appearing in the deconfinement phase is known, its effect on the glueball correlation function at finite temperature
has not been considered in previous studies.
The present study reveals that the constant contribution to the glueball correlation function
actually occurs in the three lowest-lying glueball states,
corresponding to the $0^{++}$, $2^{++}$, and $0^{-+}$ glueballs, from near the critical temperature of the deconfinement phase transition $T_C$  
to the high temperature side.
If the existence of constant terms is taken into account in the standard pole-mass analysis, 
it is observed that the pole-mass of the glueball ground state remains unchanged below $T_C$ and then 
increases with temperature above $T_C$. The result indicates that the true temperature dependence 
of the glueball mass above $T_C$ is opposite to the results of the previous studies. 
\end{abstract}

\maketitle

 
\section{Introduction}
\label{sec:INTRO}
One of the interesting topics in finite-temperature QCD is the possible mass shift and the width broadening
of hadrons. In particular, the mass modification due to deconfinement and chiral phase transitions has been the subject of many related studies, because it is related to the mass origin of hadrons. 
In particular, since in pure $SU(3)$ Yang-Mills theory the deconfinement phase transition becomes a first-order phase transition, we expect to clarify the direct relationship between the hadron mass change and the deconfinement phase transition by studying the thermal mass shifts of glueballs, which are well established as the only hadronic excitations at zero temperature in the theory~\cite{Morningstar:1999rf}. In previous studies~\cite{{Ishii:2001zq},{Ishii:2002ww},{Meng:2009hh}}, lattice QCD calculations reported that the mass of glueballs ``decreases'' with increasing temperature $T$ from near the phase transition temperature $T_C$.

\begin{figure}[h]
\includegraphics[width=1.0\linewidth,bb=0 0 1024 768,clip]{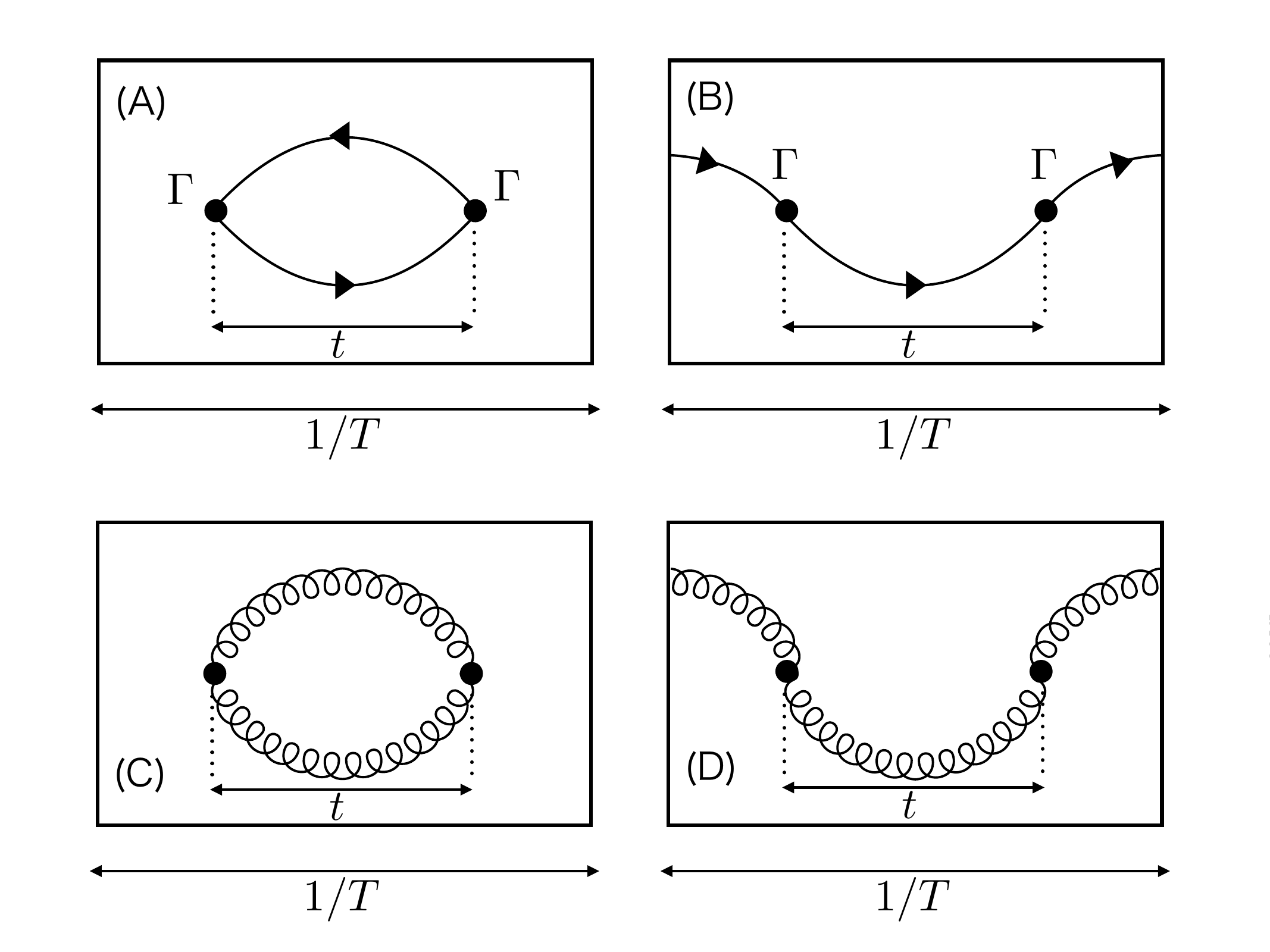}
 \caption{Schematic diagrams for a meson-like correlation (A), (B)
 and a glueball-like correlation (C), (D) in a system with period $N_t=1/(a_t T)$. The quark winding (B) and the gluon winding (D) in the time direction due to the deconfinement may lead to the constant contributions in hadron two-point correlations $C(t)$ above $T_C$. 
\label{fig:const_graph}}
\end{figure}

On the other hand, in Ref.~\cite{Umeda:2007hy}, it was pointed out that the direct effect of the deconfinement phase transition appears as a constant term at $T\gtrsim T_C$ in the calculation of the two-point correlation function of mesons, which are ordinary hadrons. This is intuitively understood to be caused by the contribution of quarks wrap rounding in the temporal direction, as shown in Fig.~\ref{fig:const_graph} (B), after the phase transition, since quarks are liberated from the confinement. 
Unless the Dirac matrix $\Gamma$ in the bilinear current operator $\bar{q}\Gamma q$ for mesons satisfies $\{\Gamma,\gamma_4\}=0$~\cite{Ohno:2011zc}, 
a constant contribution of such a quark winding appears in the time direction for mesons with zero total momentum.
Thus, there is no constant term for pseudoscalar mesons ($\Gamma=\gamma_5$) or vector mesons ($\Gamma=\gamma_i$), but it appears for scalar mesons ($\Gamma=1$) or axial-vector mesons ($\Gamma=\gamma_5\gamma_i$)~\cite{{Umeda:2007hy},{Ohno:2011zc}}.

Since the glueball with the charge-conjugation parity $C=+1$ contains two gluons as the leading Fock state~\cite{Mathieu:2008bf}, simply replacing the quark lines by gluons as in the lower panels of Fig.~\ref{fig:const_graph} will lead to a contribution from the gluon winding in the time direction due to the deconfinement, as shown in Fig.~\ref{fig:const_graph} (D). 
Therefore, the constant term is expected to appear universally in the $C=+1$ glueball two-point correlation functions, independent of the other quanta, such as the total spin $J$ and the parity $P$, without any restrictions due to the nature of the Dirac spinor, as was the case for mesons~\cite{{Umeda:2007hy},{Ohno:2011zc}}. 
Furthermore, the gluon winding contribution as shown in Fig.~\ref{fig:const_graph} (D) has a $Z_3$ center symmetry
as well as Figs.~\ref{fig:const_graph} (A) and (C), while
the quark winding contribution as shown in Fig.~\ref{fig:const_graph} (B) is not invariant under the $Z_3$ transformation.

As the previous studies~\cite{{Ishii:2001zq},{Ishii:2002ww}} 
were earlier than Ref.~\cite{Umeda:2007hy}, the analysis did not take into account the presence of the constant terms that
are likely to appear in the low-lying glueball two-point functions above $T_C$ as well. Instead, in Refs.~\cite{{Ishii:2002ww},{Meng:2009hh}}, the glueball two-point functions are analyzed under the assumption that the spectral function may be affected by the thermal width.

In this study, we first attempt to confirm that constant terms do indeed appear in the glueball two-point functions with respect to the deconfinement phase transition, and then will perform the standard pole-mass analysis to determine the glueball ground-state masses at finite temperature, taking into account the presence of such contributions. Recall that three lowest-lying glueball 
states ($J^{PC}=0^{++}$, $2^{++}$, and $0^{-+}$) have the $C$-parity $C=+1$~\cite{Morningstar:1999rf}.

This paper is organized as follows. 
In Sec.~\ref{sec:SETUP}, we first briefly summarize
our numerical simulations for the glueball spectroscopy at zero temperature using anisotropic lattice QCD. 
Section~\ref{sec:NUM_RESULTS} presents the numerical results of the glueball two-point functions at the finite temperature including the deconfinement phase above the critical temperature $T_C$. We verify the presence of constant terms in the glueball two-point functions above $T_C$ and then examine the true temperature dependence of the glueball ground-state masses for the three lowest-lying channels ($J^{PC}=0^{++}$, $2^{++}$, and $0^{-+}$) above $T_C$ with the constant term subtracted. Finally, we close with a summary in Sec.~\ref{sec:SUMMARY}.

%
%
%

\section{Lattice setup}
\label{sec:SETUP}

\subsection{Gauge ensembles}
We perform the pure Yang-Mills lattice simulations
on anisotropic lattices
using the anisotropic Wilson gauge action
\begin{align}
S_W[U]
&=
\frac{1}{\xi_0}S_{\mathrm{sp}}[U]+
\xi_0 S_{\mathrm{tp}}[U],
\end{align}
where the parameter $\xi_0$ is the bare anisotropy~\cite{Klassen:1998ua}. The explicit expressions of 
the spatial plaquette (sp) action
$S_{\mathrm{sp}}[U]$ and 
the temporal plaquette (tp) action
$S_{\mathrm{tp}}[U]$ are given by
\begin{align}
S_{\mathrm{sp}}[U]&=\frac{\beta}{3}\sum_{i<j\le 3}
{\rm Re}{\rm Tr}\left[1-P_{ij}(x)
\right], \\
S_{\mathrm{tp}}[U]&=\frac{\beta}{3}\sum_{i\le 3}
{\rm Re}{\rm Tr}\left[
1-P_{i4}(x)
\right],
\end{align}
where $\beta$ is associated with the bare QCD coupling constant $g_0$ as $\beta=6/g_0^2$ and $P_{\mu \nu}(x)$ 
denotes the plaquette operator defined as
\begin{align}
P_{\mu \nu}(x)&=U_\mu(x)U_\nu(x+\mu)U^{\dagger}_\mu(x+\nu)U^{\dagger}_\nu(x)
\end{align}
in the $\mu$-$\nu$ plane. 
If $\xi_0 > 1$ is chosen, the temporal lattice spacing $a_t$ is finer than the spatial lattice spacing $a_s$. Since the true anisotropy $\xi\equiv a_s/a_t$ 
is equal to $\xi_0$ only at the classical level, one needs to determine the renormalized anisotropy $\xi$ as a function of the
bare anisotropy $\xi_0$ at a given $\beta$~\cite{Klassen:1998ua}.
In this study, we adopt the parameter set of $\beta=6.25$ and $\xi_0=3.2552$, which
yields the renormalized anisotropy $\xi=4$~\cite{Klassen:1998ua}, following the simulations performed in Refs.~\cite{{Ishii:2001zq},{Ishii:2002ww}}.

We generate anisotropic pure gauge ensemble with lattice size of $N_s^3\times N_t$ using the anisotropic Wilson gauge action at $\beta=6.25$ and $\xi_0=3.2552$. The gauge configurations are separated by 200 sweeps after 5000 sweeps for thermalization in all simulations. Each sweep consists of one pseudo heat-bath \cite{Cabibbo:1982zn} combined with four over-relaxation \cite{Creutz:1987xi} steps. 
For the ``zero-temperature'' simulation ($N_s^3\times N_t=20^3\times 80$), 4000 gauge configurations are totally accumulated. Details of the Wilson gauge action adopted on our anisotropic lattice at zero temperature are shown in Table~\ref{tab:gauge_ensemble}.

\subsection{Glueball spectroscopy at zero temperature}

For the glueball spectroscopy, we will calculate four lowest-lying glueball states which carry quantum numbers $R^{PC}=A_1^{++}$, $A_1^{-+}$, $E^{++}$ and
$T_2^{++}$, where the irreducible representations (irreps) $R$ of the octahedral point group $O_h$ are the counterparts of the spin $J$ for the continuum rotational group, and $P$ and $C$ represent the parity and the charge-conjugation parity, respectively.
The quantum number $R^{PC}$ is expected
to have the following correspondence:
$0^{++}\leftrightarrow A_1^{++}$, $0^{-+}\leftrightarrow A_1^{-+}$, and $2^{++}\leftrightarrow E_1^{++}\oplus T_2^{++}$ in the continuum limit~\cite{{Johnson:1982yq},{Berg:1982kp}}. 

\begin{figure}[h]
\includegraphics[width=0.6\linewidth,bb=0 0 318 241,clip]{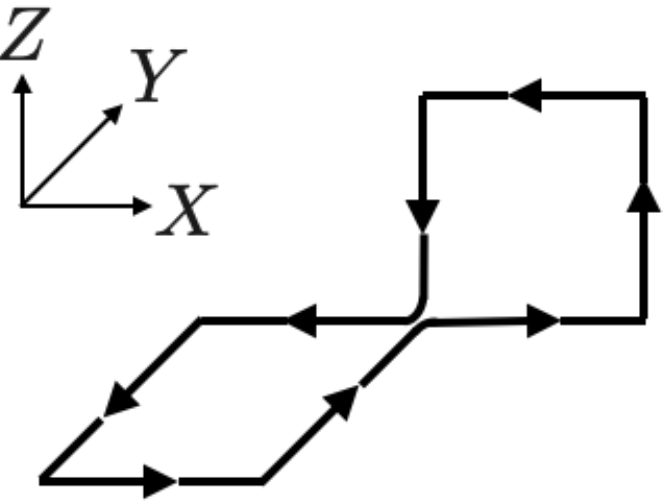}
 \caption{Our choice of a ``fish-shaped'' Wilson loop with eight links for the construction of the glueball operator in this study.
\label{fig:glueball_op_fish}}
\end{figure}

We choose a ``fish-shaped'' operator (denoted as ${\cal O}_{\mathrm{GB}}$) , {\it i.e.}, spacelike Wilson loops of eight links as shown in Fig.~\ref{fig:glueball_op_fish}, to construct the glueball operators for $R^{PC}=A_1^{++}$, $A_1^{-+}$, $E^{++}$, and $T_2^{++}$~\footnote{In Ref.~\cite{Sakai:2022zdc}, after applying high diffusion by the stout-link smearing method, it was found that the shape dependence of the Wilson loop almost disappears in the resulting two-point function.}. We then get
the irreducible contents of the representation $R^P$ with fixed $C$-parity from the operator ${\cal O}_{\mathrm{GB}}$ according to Ref.~\cite{Berg:1982kp}. 

%
%
\begin{table*}[ht]
\caption{
Details of the simulation parameters for the Wilson action adopted on our anisotropic lattice, following the simulations performed in Refs.~\cite{{Ishii:2001zq},{Ishii:2002ww}}.
\label{tab:gauge_ensemble}}
\begin{ruledtabular} 
\begin{tabular}{lccccccc}
\hline
 $\beta$ & $N_s^3 \times N_t$ & $\xi_0$ & $\xi$ & $a_s^{-1}$ [GeV] & ($a_s$ [fm]) &  $a_t^{-1}$ [GeV] & ($a_t$ [fm]) 
 \cr
 \hline
 6.25 & $20^3 \times 80$ & 3.2552 & 4 & 2.341(16) &($\sim$ 0.084) & 9.365(66)& ($\sim$ 0.021)  
\cr \hline
\end{tabular}
\end{ruledtabular}
\end{table*}
%

%
%

The operator ${\cal O}_{\mathrm{GB}}$ is classified with the shape characterized by the ordered closed path ${\cal C}$ and its associated orientation $[X,Y,X,Z,-X,-Z,-X,-Y]$. 
A linear combination of 48 copies with different orientations of Wilson loops 
${\cal O}_{\mathrm{GB}}[\hat{S}_i{\cal C}]$ where 48 symmetry operators
$\hat{S}_i$ are applied to the original
orientation is performed as
\begin{align}
{\cal O}_{\mathrm{GB}}^{\Gamma}=\hat{P}_\Gamma{\cal O}_{\mathrm{GB}}[{\cal C}]
=\frac{1}{48}\sum_{i=1}^{48}\chi^{\ast}_{\Gamma}(S_i){\cal O}_{\mathrm{GB}}[\hat{S}_i{\cal C}],
\end{align}
where $\chi_{\Gamma}(S_i)$ denotes
the characters of the irreps $\Gamma=A_1^{+}$, $A_1^{-}$, $E^{+}$, and $T_2^{+}$. 
We then construct two-point correlation functions of the glueball states for given irreps $\Gamma$ as
\begin{align}
C(t)=\sum_{t^\prime}\langle 0|
\widetilde{\cal O}_{\mathrm{GB}}^\Gamma(t+t^\prime)
\widetilde{\cal O}_{\mathrm{GB}}^\Gamma(t^\prime)^\dagger|0\rangle,
\end{align}
where the tilde over ${\cal O}_{\mathrm{GB}}$ implies the vacuum-subtracted operator defined as $\widetilde{\cal O}_{\mathrm{GB}}^\Gamma={\cal O}_{\mathrm{GB}}^\Gamma -\langle 0|{\cal O}_{\mathrm{GB}}^\Gamma|0\rangle$. See Ref.~\cite{Sakai:2022zdc} for further details.

As is well known, ultraviolet noise from the gauge fields makes it difficult to perform the glueball mass calculation without noise reduction technique. In this study, we adopt the stout-link smearing method for the construction of the extended glueball operators which have a better overlap with the glueball ground states~\cite{Sakai:2022zdc}.

The stout-link smearing is a well-established smearing scheme and is defined as the following recursive procedure~\cite{Morningstar:2003gk}. 
In this work, for the stout-smearing parameters, the spatially isotropic three-dimensional parameter set is chosen
to be $\rho_{ij}=\rho$ and $\rho_{4\mu}=\rho_{\mu 4}=0$. 
Therefore, only the spatial link variables
$U_i^{(k)}(x)$ at step $k$ are mapped into the link variables $U_i^{(k+1)}(x)$ as below
\begin{align}
U_i^{(k+1)}(x)=\exp\left(
i\rho Q_i^{(k)}(x)
\right)U_i^{(k)}(x),
\end{align}
where $Q_i^{(k)}(x)$ corresponds to
a Lie algebra valued quantity given by
\begin{align}
Q_i^{(k)}(x)=ig_0^2\partial_{x,i}S_{\mathrm{sp}}[U^{(k)}]
\label{eq:flow_rep}
\end{align}
with the spatial plaquette action $S_{\mathrm{sp}}$ in terms of the stout-link variables~\footnote{For 
an isotropic four-dimensional parameter set as $\rho_{\mu\nu}=\rho$ with the case of $\xi_0=1$, $Q_\mu^{(k)}$ can be expressed by $Q_\mu^{(k)}=ig_0^2\partial_{x,\mu}S_W[U^{(k)}]$, which was first derived in Ref.~\cite{Luscher:2010iy}, and its particular form has recently led to a direct analytical proof of the equivalence between stout-link smearing and Wilson flow~\cite{Nagatsuka:2023jos}.
}.
The operator $\partial_{x,\mu}$ stands for the Lie-algebra valued differential operator with respect to the link variable~\cite{Luscher:2010iy}.

In this study, we use $\rho=0.1$ for the smearing parameter. 
It is worth mentioning that when the number of smearing steps is $N_{\mathrm{smr}}$, the spatial diffusion radius (denoted as $R_d$) resulting from the smearing process is approximately given by $R_d=\sqrt{6\rho N_{\mathrm{smr}}}$~\cite{{Sakai:2022zdc},{Nagatsuka:2023jos}}.

After applying the stout-link smearing, the glueball two-point function can be dominated by the ground state (its mass and spectral weight, denoted as $M$ and $W$, respectively) and then
expressed as $C(t) \sim W e^{-Mt}$ for the relatively large Euclidean time $t$. However, in the time direction, 
the periodic boundary condition (p.b.c.) is imposed with the finite extent $N_t$ so that the two-point function is supposed to behave as below 
\begin{widetext}
\begin{align}
C(t)\xrightarrow[\mathrm{p.b.c.}]{t/a\gg 1} \sum_{n=-\infty}^{\infty}We^{-M (t+nN_t)}
&=\frac{W}{1-e^{-M N_t}}\left(
e^{-Mt}+e^{-M(N_t-t)}\right)\\
&=W\frac{\cosh[M(t-N_t/2)]}{\sinh[M N_t/2]}, 
\end{align}
\end{widetext}
where the ``wrap-around effect'' 
across the time boundary can be represented as $\sum_{n=-\infty}^\infty C(t+nN_t)$ (see Ref.~\cite{Sasaki:2005ug} for more details). 
Therefore, to properly account for the wrap-around effect
we use the effective mass $M_{\mathrm{eff}}(t)$
that is defined as a solution of 
\begin{align}
\frac{C(t)}{C(t+1)}=\frac{\cosh[M_{\rm eff}(t)(t-N_t/2)]}
{\cosh[M_{\rm eff}(t)(t+1-N_t/2)]
}.
\label{eq:effmass_cosh}
\end{align}

\begin{figure}[t]
\includegraphics[width=0.48\linewidth,bb=0 0 792 612,clip]{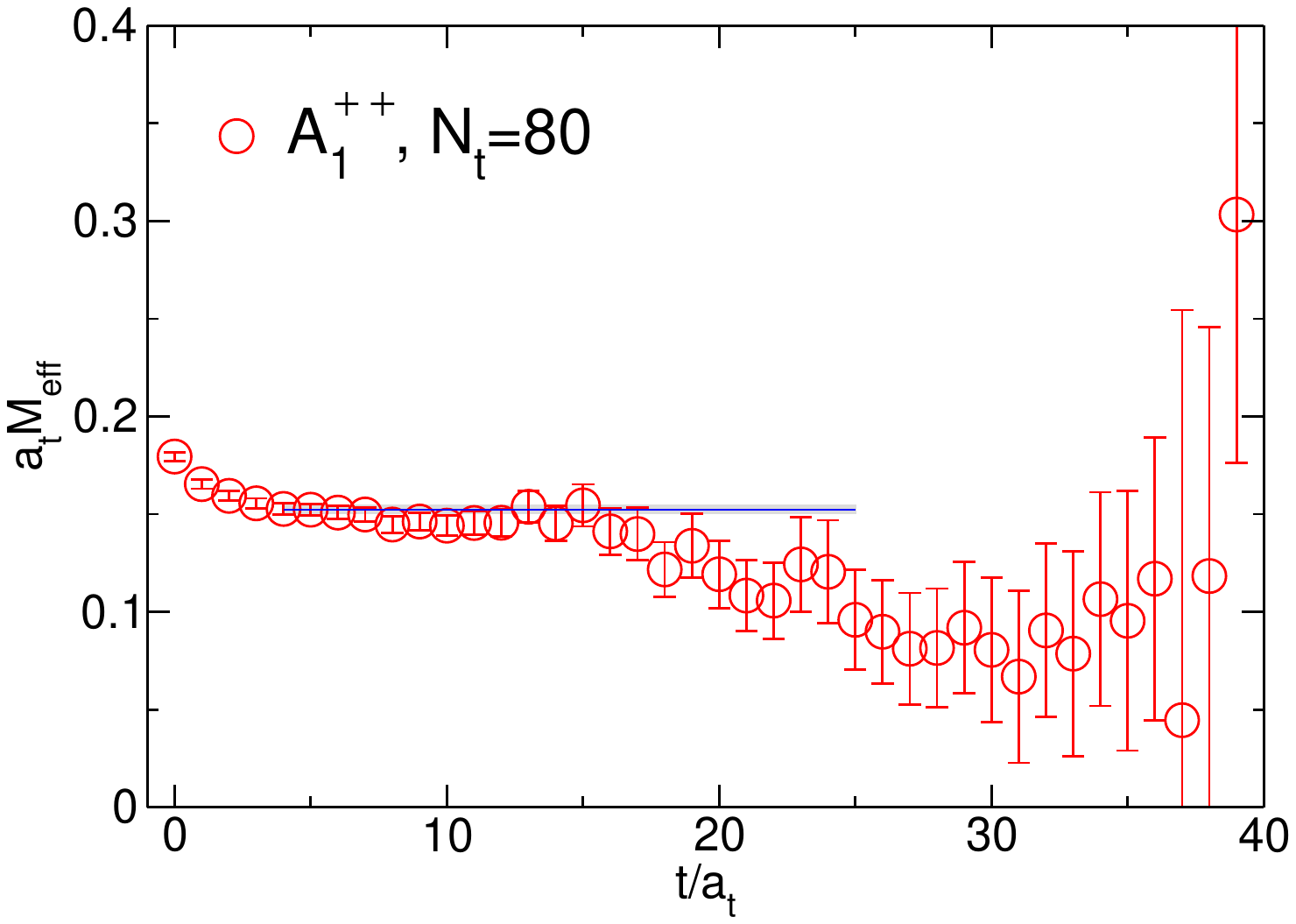}
\includegraphics[width=0.48\linewidth,bb=0 0 792 612,clip]{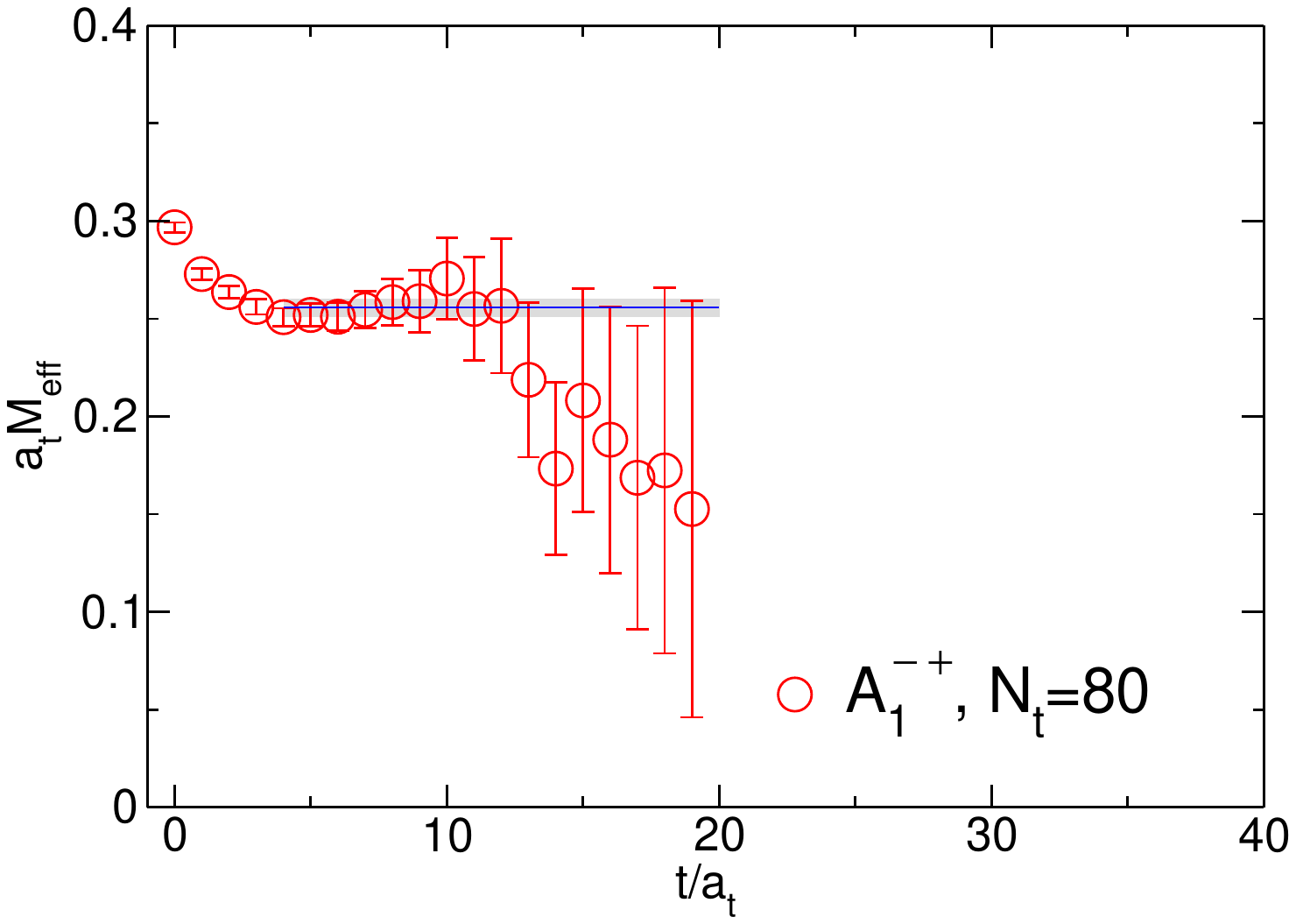}
\includegraphics[width=0.48\linewidth,bb=0 0 792 612,clip]{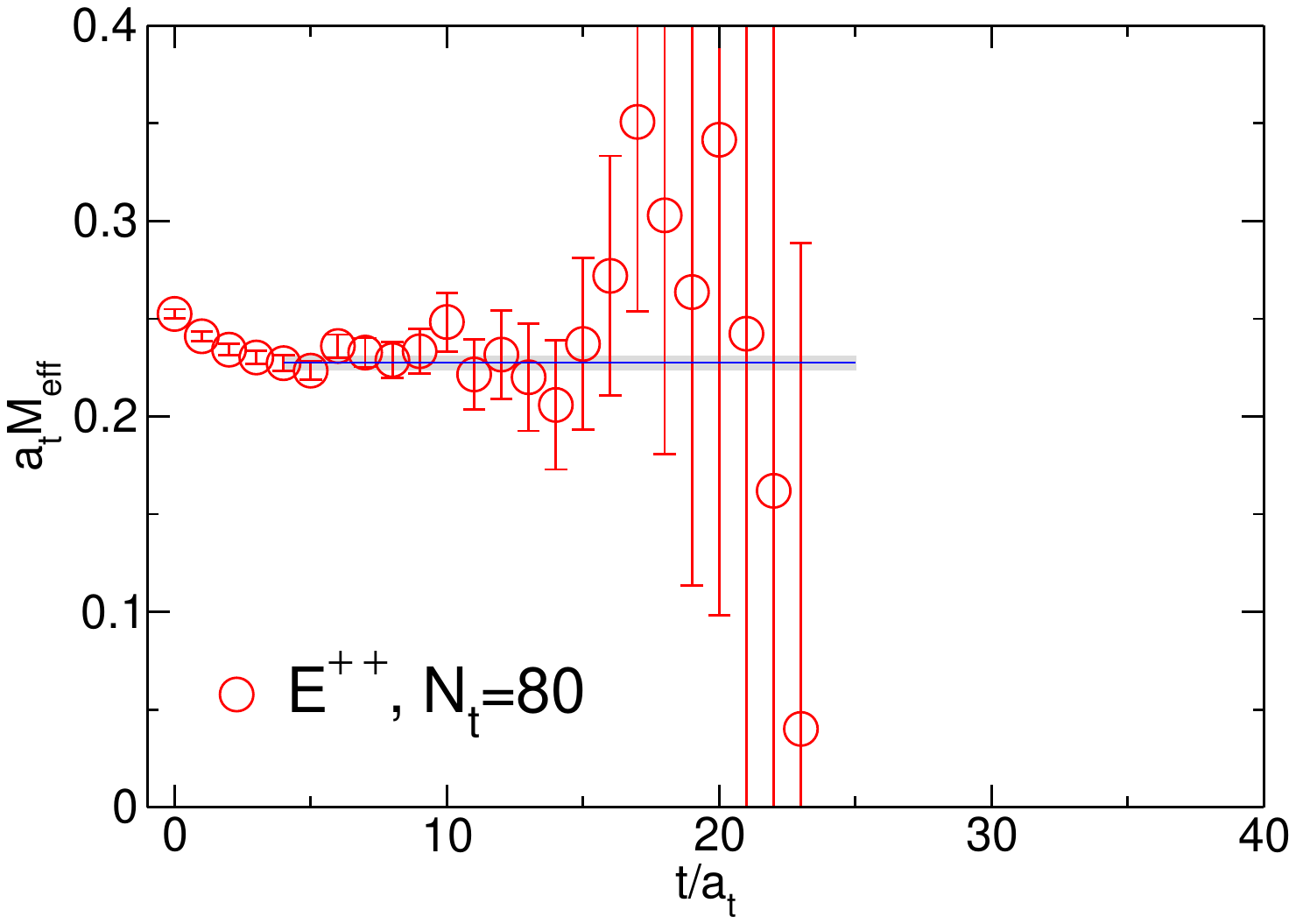}
\includegraphics[width=0.48\linewidth,bb=0 0 792 612,clip]{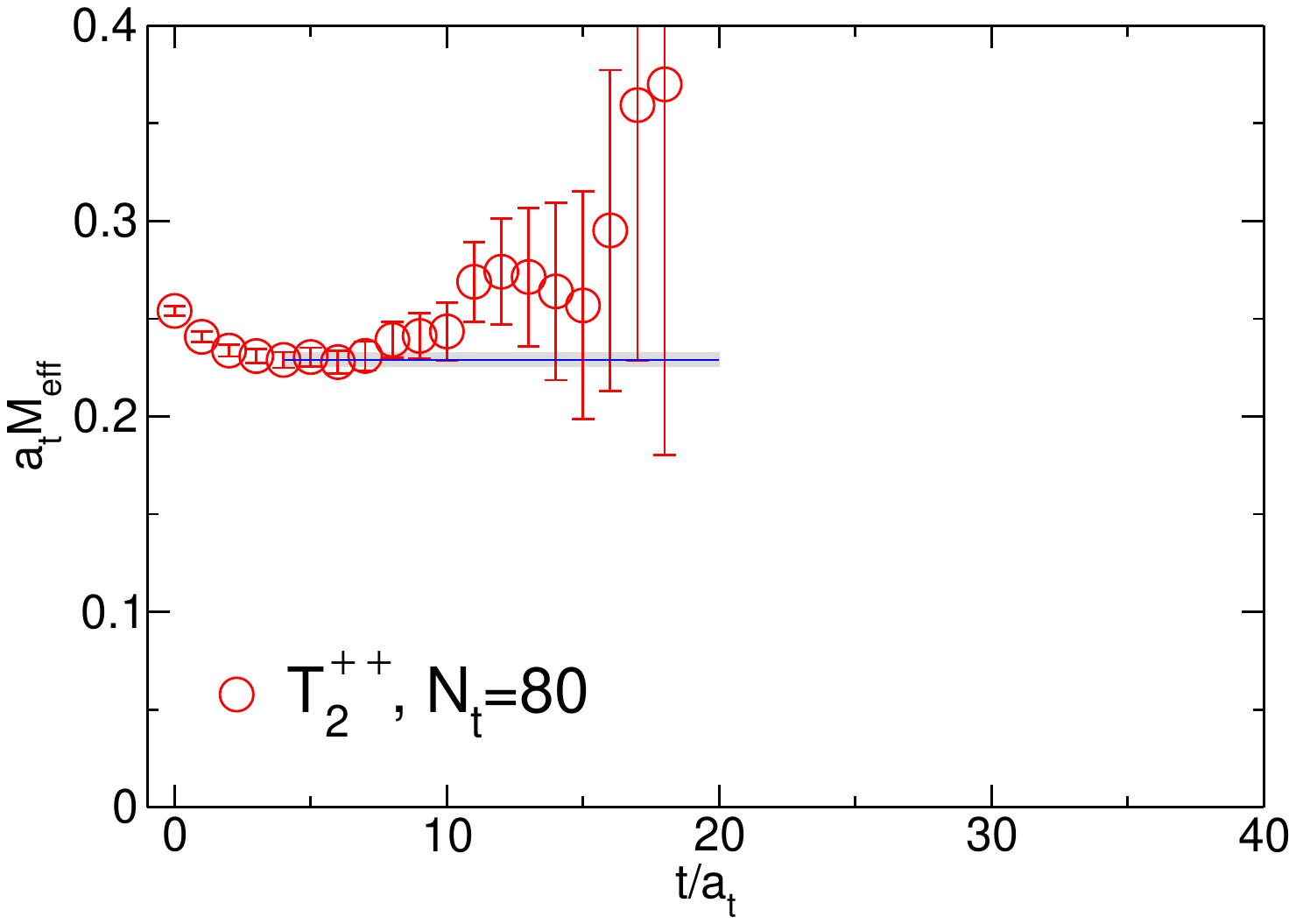}
 \caption{
Effective mass plots for $A_1^{++}$ (upper left panel), $A_1^{-+}$ (upper right panel),
$E^{++}$ (lower left panel) and $T_2^{++}$ (lower right panel) irreps with $N_{\rm smr}=70$.
Solid lines indicate fit results and shaded bands display the fitting range and one standard
deviation estimated by the jackknife method.
\label{fig:Mg_zero_temp}}
\end{figure}

Figure~\ref{fig:Mg_zero_temp} shows the effective masses for $A_1^{++}$ (upper-left panel), $A_1^{-+}$ (upper-right panel), $E^{++}$ (lower-left panel) and $T_2^{++}$ (lower-right panel) irreps with $N_{\rm smr}=70$. 
In each panel of this figure, the horizontal solid lines represent each fit result obtained by a correlated fit using a single-cosh function form, $A \cosh[M_G(t-N_t/2)]$ with
two parameters of $A$ and $M_G$ on $\widetilde{C}(t)\equiv C(t)/C(0)$, and
shaded bands display the fit range and one standard deviation.
In Table~\ref{Table:Mg_zero_temp}, we summarize the results of the ground-state mass in all four channels.

\begin{figure}[ht]
\includegraphics[width=0.48\linewidth,bb=0 0 792 612,clip]{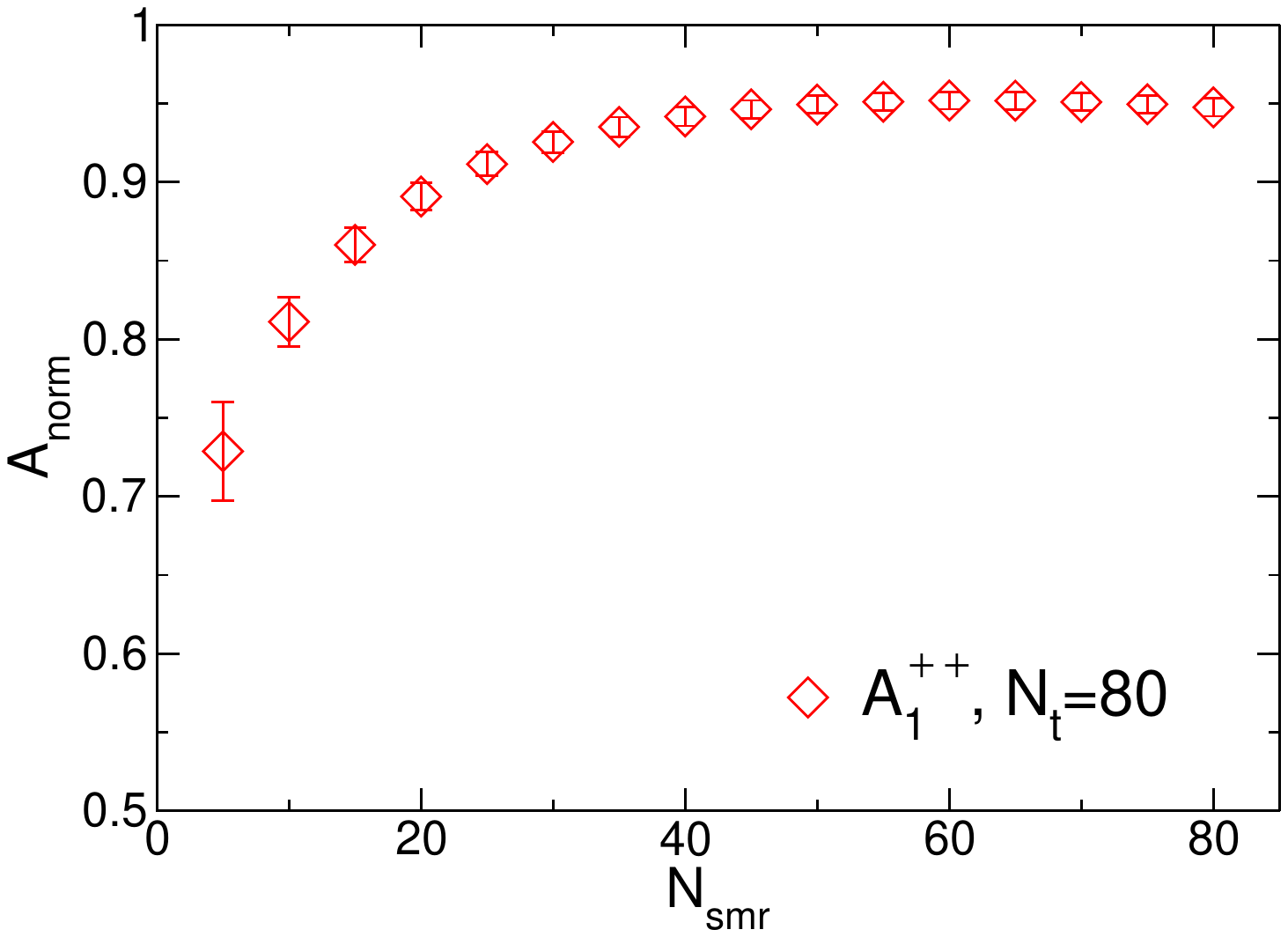}
\includegraphics[width=0.48\linewidth,bb=0 0 792 612,clip]{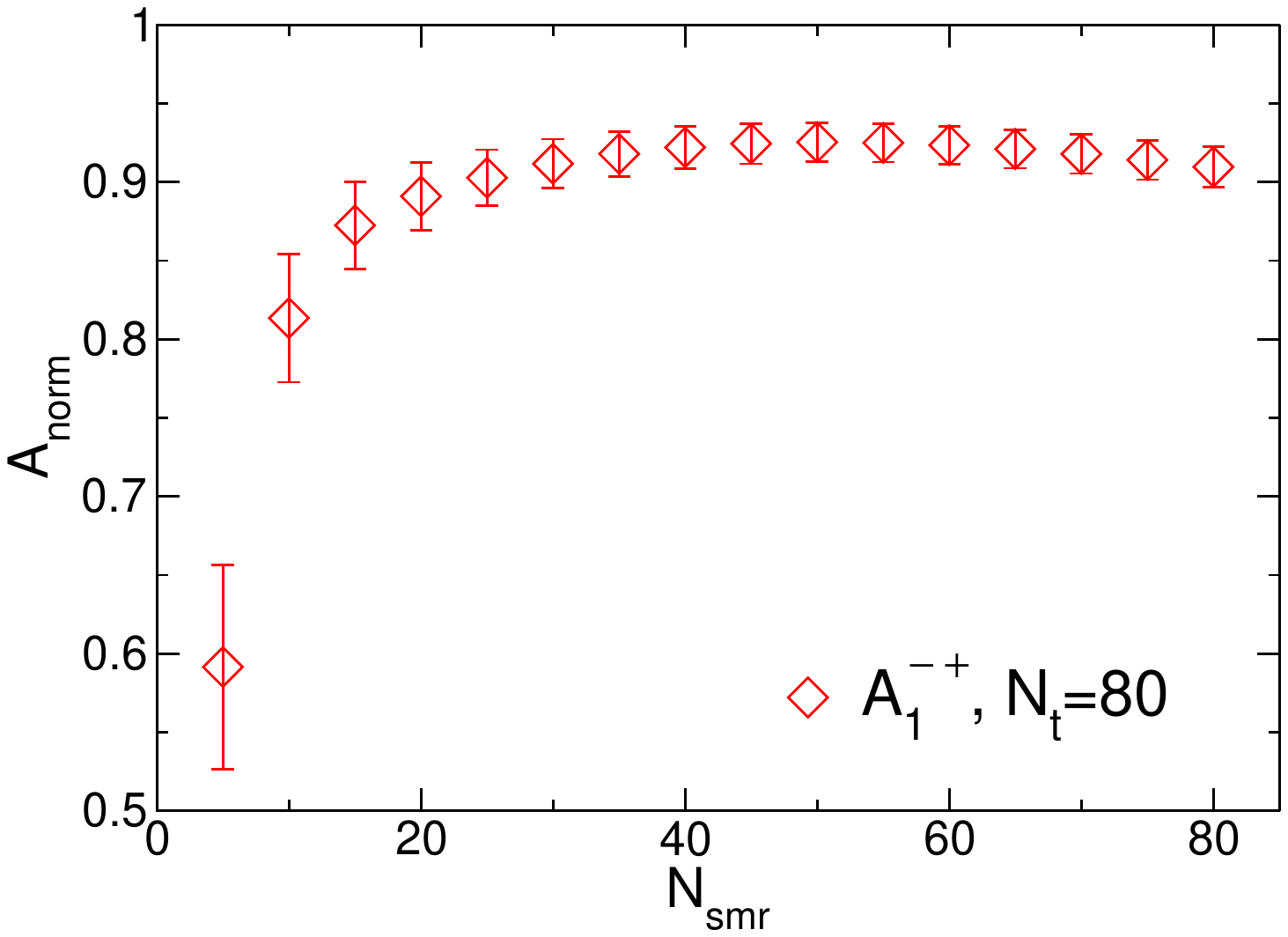}
\includegraphics[width=0.48\linewidth,bb=0 0 792 612,clip]{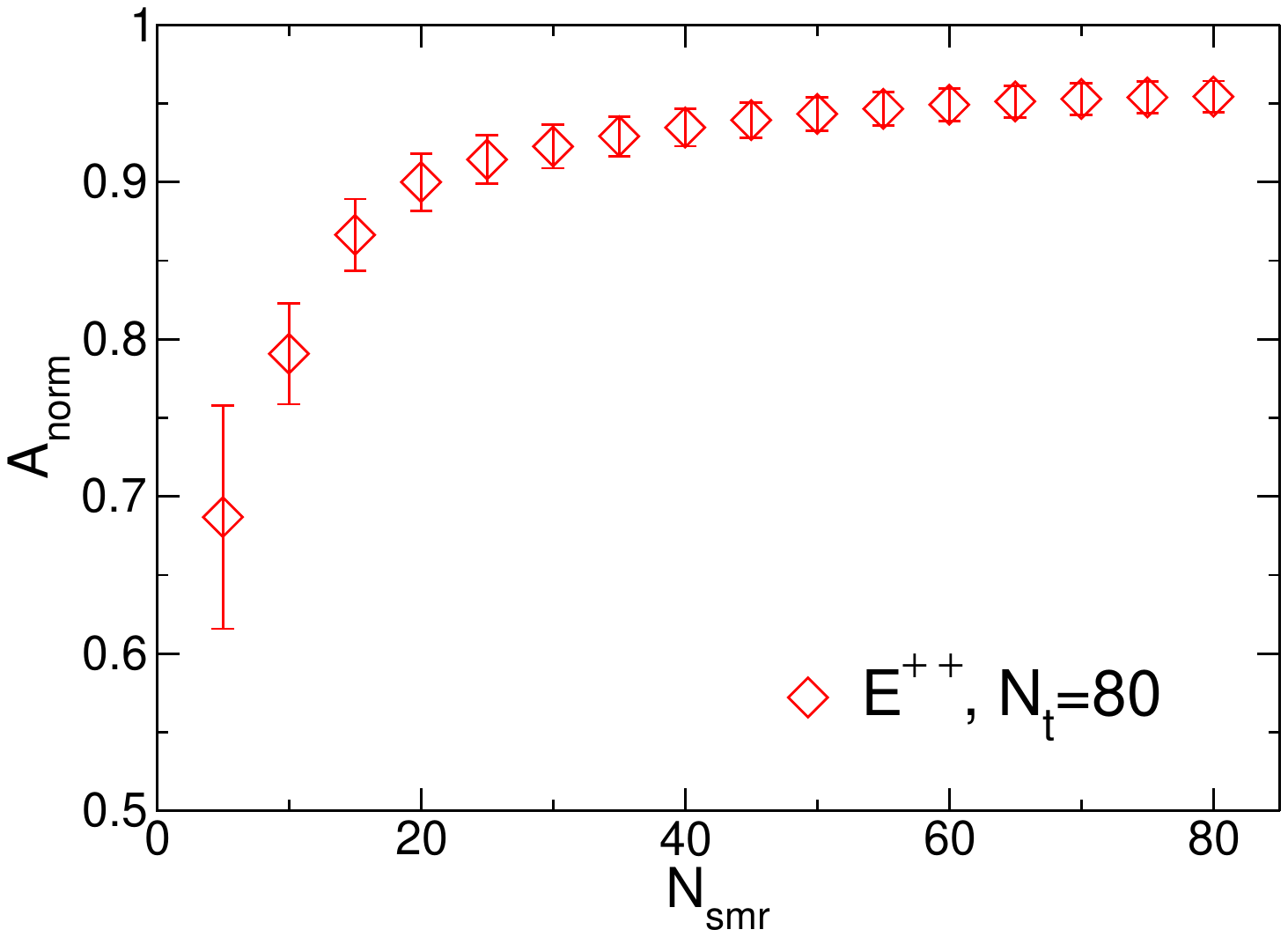}
\includegraphics[width=0.48\linewidth,bb=0 0 792 612,clip]{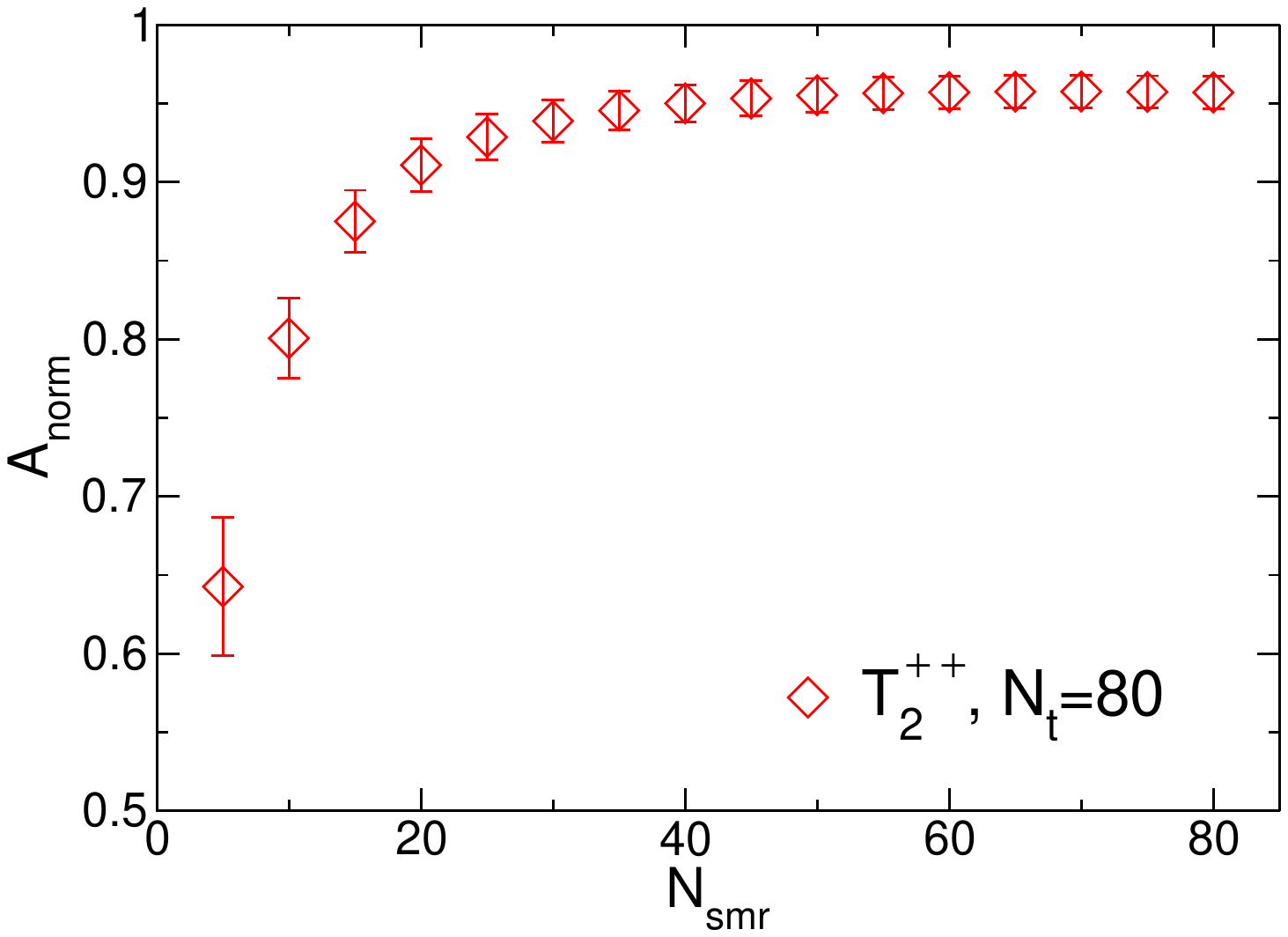}
 \caption{
Normalized amplitude, $A_{\rm norm}=A\cosh[M_GN_t/2]$, for $A_1^{++}$ (upper left panel), $A_1^{-+}$ (upper right panel),
$E^{++}$ (lower left panel) and $T_2^{++}$ (lower right panel) irreps 
as a function of $N_{\rm smr}$.
\label{fig:Ampl_zero_temp}}
\end{figure}

The amplitude $A$ is not normalized due to the wrap-round effect even if the two-point function is solely described by the ground-state contribution. Therefore, we define the normalized amplitude $A_{\mathrm{norm}}=A\cosh[M_G N_t/2]$, which is defined such that $0\le A_{\mathrm{norm}} \le 1$ regardless of
the size of $N_t$. In Fig.~\ref{fig:Ampl_zero_temp}, we show the normalized amplitude $A_{\rm norm}$ for $A_1^{++}$ (upper left panel), $A_1^{-+}$ (upper right panel),
$E^{++}$ (lower left panel) and $T_2^{++}$ (lower right panel) irreps 
as a function of $N_{\rm smr}$. In all four channels, the magnitude of the normalized amplitude is well saturated near unity ($A_{\mathrm{norm}}\sim 0.9$-0.95) around $N_{\rm smr}=70$, where the fit mass does not change within the statistical uncertainty as $N_{\rm smr}$ is varied.
The stout-link smearing method has been demonstrated to produce highly stable results with respect to the smearing number $N_{\rm smr}$. The resulting two-point function composed of smeared operators is efficiently dominated by the contribution of the ground state.

Therefore, in the following analysis, we simply set $N_{\mathrm{smr}}=70$, which corresponds to the spatial diffusion radius, $R_d\approx 0.54$ fm.

%
%
%

%
%
%

%
%
\begin{table}[hb]
\centering
\caption{
Masses of the ground state of the $A_1^{++}$, $E^{++}$, $T_2^{++}$ 
and $A_1^{-+}$ irreps with $N_{\rm smr}=70$ at zero temperature on $20^3 \times 80$ lattice.
\label{Table:Mg_zero_temp}
}
\begin{ruledtabular} 
\begin{tabular}{c c c  c c c}
\hline
Irreps & $a_t M_G$ & $M_G$ [GeV] & Fit range & $\chi^2/{\rm dof}$ \cr
\hline
$A_1^{++}$ & 0.1523(21) & 1.426(20) & [4, 25] & 1.41  \cr
$E^{++}$    &  0.2274(35) & 2.130(33) & [4, 25] & 1.55  \cr
$T_2^{++}$ &  0.2291(35) & 2.146(33)&[4, 20] & 0.73  \cr
$A_1^{-+}$ &  0.2509(42) & 2.350(39)& [4, 20] & 0.44 \cr
\hline
\end{tabular}
\end{ruledtabular}
\end{table}

\section{Numerical results at finite temperature}
\label{sec:NUM_RESULTS}

We next calculate the glueball two-point functions on the lattice sizes of $20^3\times N_t$ with $N_t=72$, 48, 40, 36, 34, 32, 28, 24, and 20, covering a wide range of $130 \le T \le 468$ [MeV], where the temperature $T$
is defined by $T=\frac{1}{a_t N_t}$.
The critical temperature $T_C$ for
the deconfinement phase transition on our anisotropic lattice was estimated as $T_C\approx 280$ MeV in the previous work~\cite{{Ishii:2001zq},{Ishii:2002ww}}.
In addition, we also use the lattice sizes of $16^3\times N_t$ and $24^3 \times N_t$ with $N_t=24$ ($T\approx 0.5T_C < T_C$) and 72 ($T\approx 1.5T_C > T_C$), which are reserved for the finite volume study. The number of configurations
analyzed is ${\cal O}(2000\mbox{-}10000)$ as shown in Table~\ref{tab:finite_T_configs}.

\subsection{Verification of the constant term}

To determine the presence of a constant term in the two-point correlation function above $T_C$,
we first define the time-derivative two-point function as
\begin{align}
D(t) &\equiv \frac{1}{2}\left(C(t+1) - C(t-1)\right),
\end{align}
which eliminates the constant term in $C(t)$ and then has the following functional form: $D(t)\propto \sinh[M_G(t-\frac{N_t}{2})]$.
For this time-derivative correlator $D(t)$, the effective mass is given by a solution of
\begin{align}
\frac{D(t)}{D(t+1)}&=\frac{\sinh[\widetilde{M}_{\rm eff}(t)(t-N_t/2)]}
{\sinh[\widetilde{M}_{\rm eff}(t)(t+1-N_t/2)]
}.
\end{align}
If it is the case of $\widetilde{M}_{\rm eff}(t)={M}_{\rm eff}(t)$, one can 
confirm that there is no constant term in the original two-point function $C(t)$.

%
%
\begin{figure}[t]
\includegraphics[width=0.48\linewidth,bb=0 0 792 612,clip]{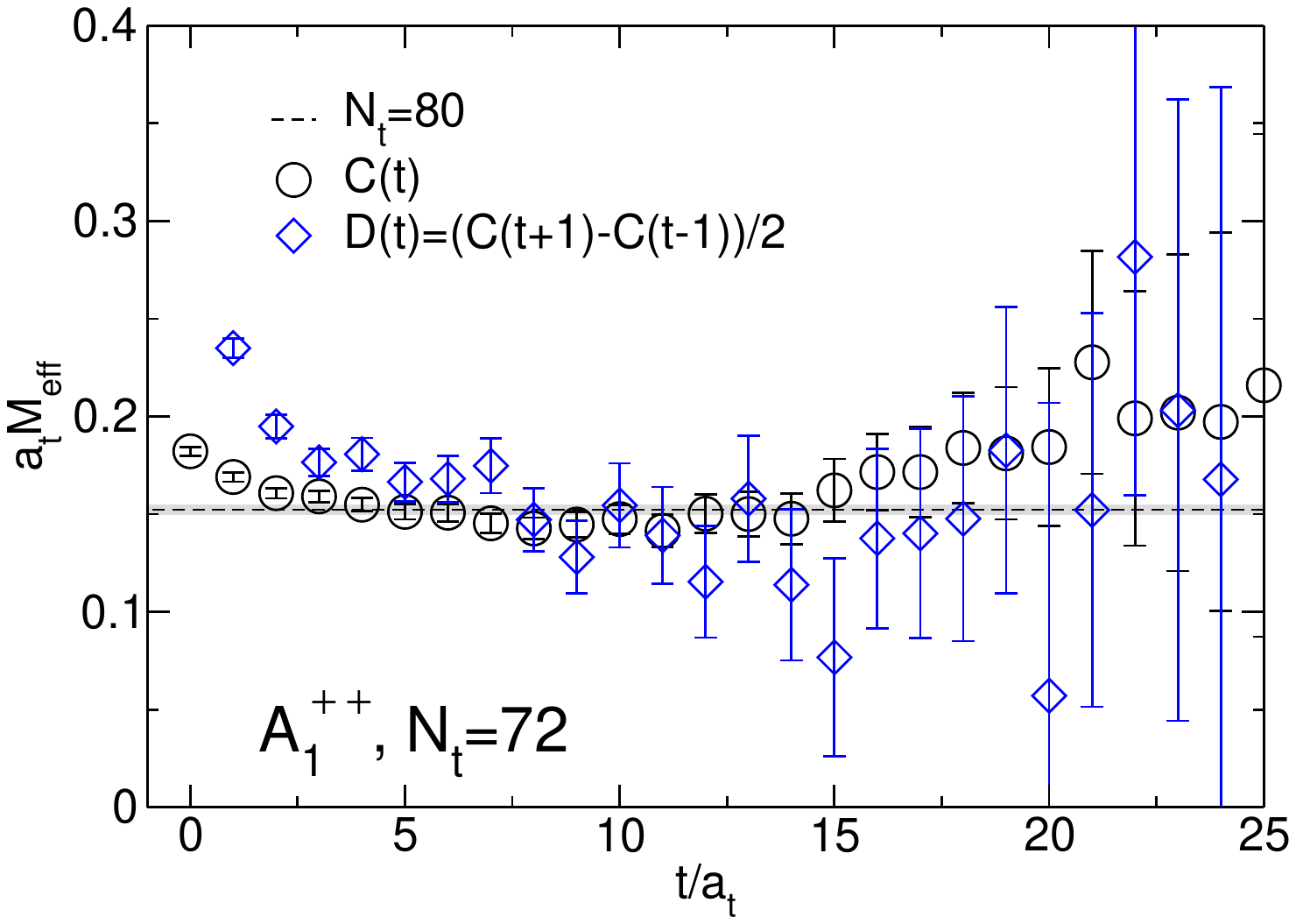}
\includegraphics[width=0.48\linewidth,bb=0 0 792 612,clip]{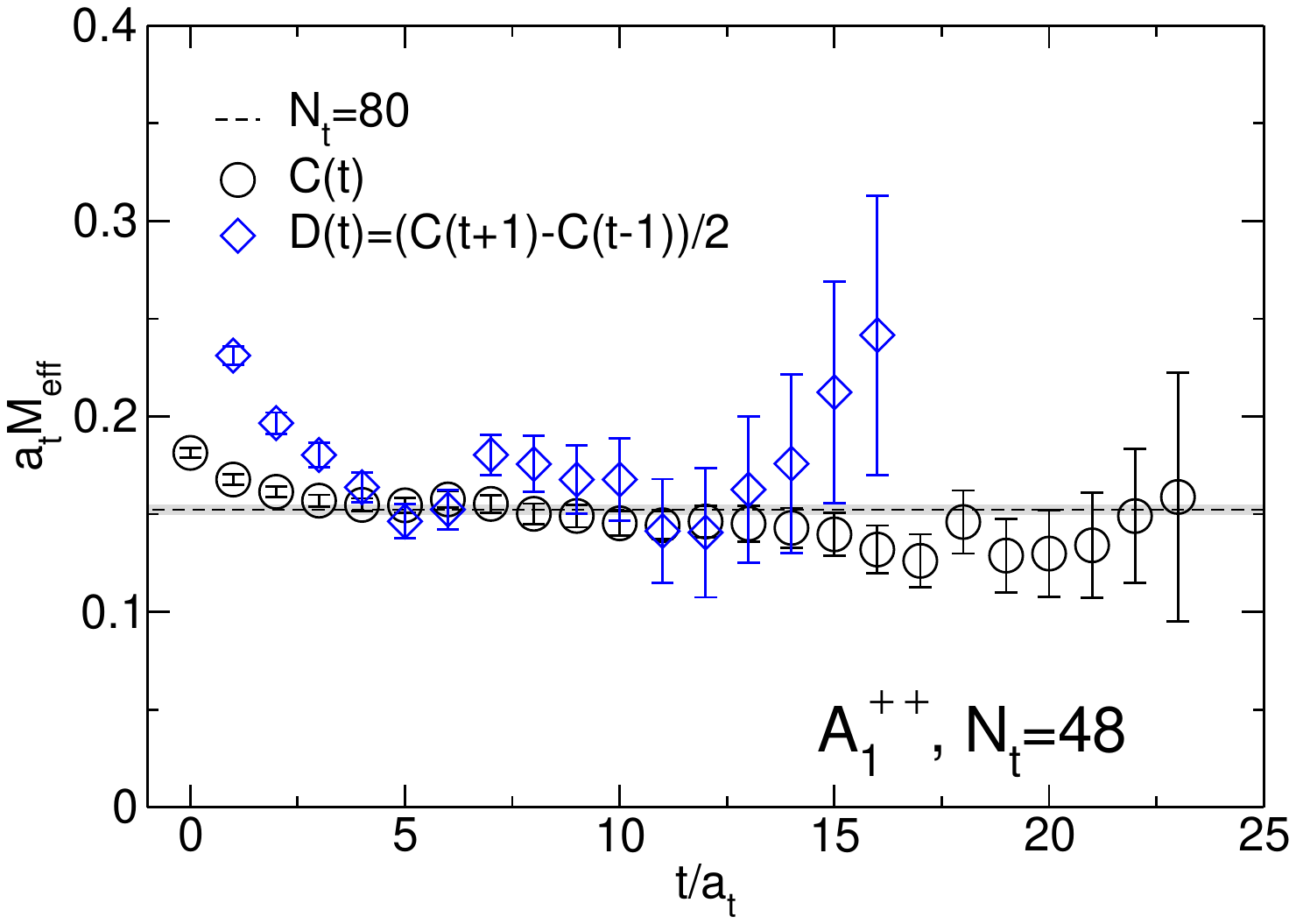}
\includegraphics[width=0.48\linewidth,bb=0 0 792 612,clip]{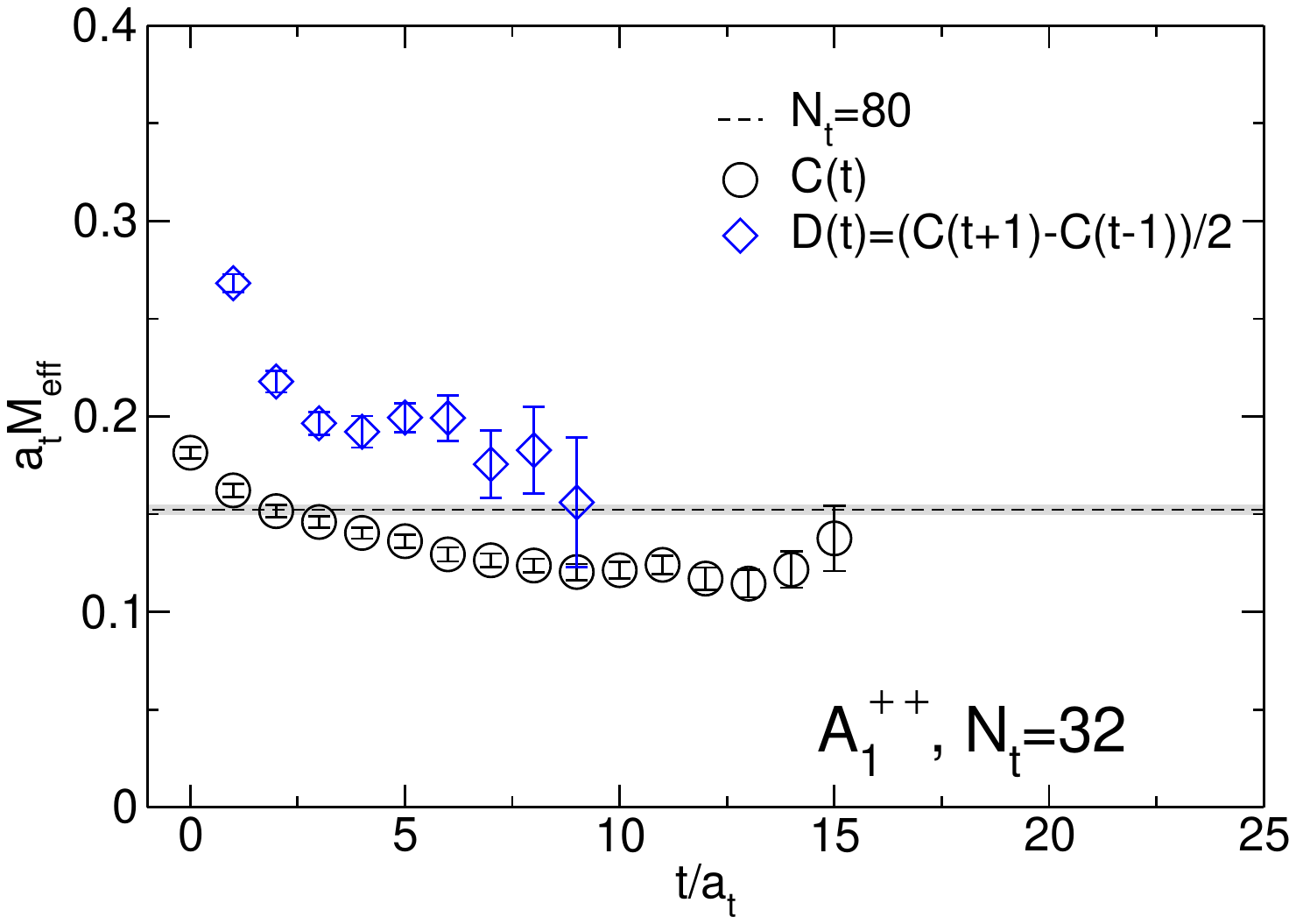}
\includegraphics[width=0.48\linewidth,bb=0 0 792 612,clip]{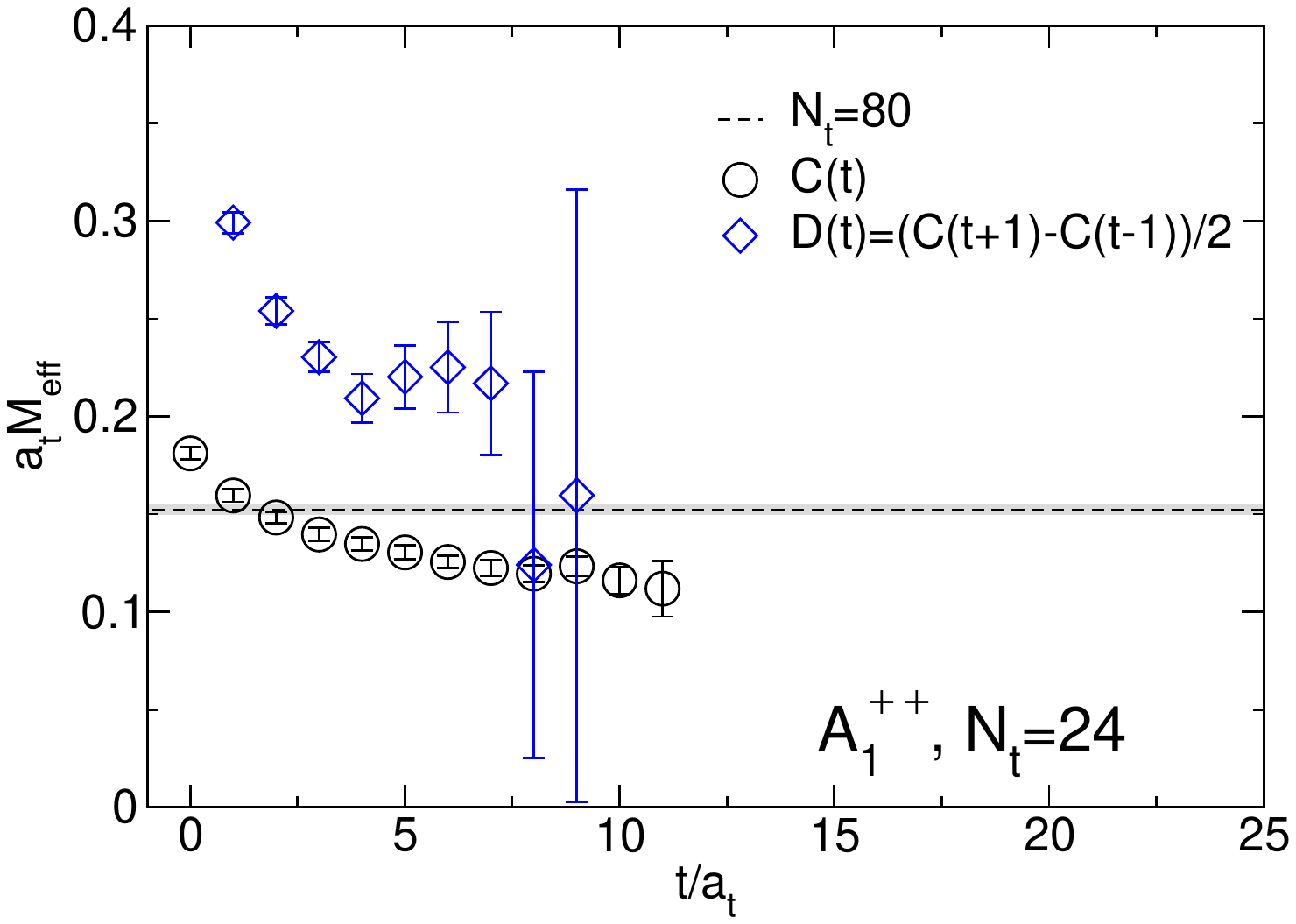}
 \caption{
Effective mass plots of the
$A_1^{++}$ irrep for $N_t=72$ (upper left panel), $N_t=48$ (upper right panel),
$N_t=32$ (lower left panel) and $N_t=24$ (lower right panel) with $N_{\rm smr}=70$.
Black circle and blue diamond symbols are given by the two-point correlator $C(t)$ and 
its time-derivative one $D(t)=(C(t+1)-C(t-1))/2$. 
Dotted lines with shaded bands indicate the fit result for $N_t=80$ as $T=0$.
\label{fig:Mg_comp}}
\end{figure}

Figure~\ref{fig:Mg_comp} shows effective mass plots of the $A_1^{++}$ irrep for $N_t=72$ (upper left panel), $N_t=48$ (upper right panel),
$N_t=32$ (lower left panel) and $N_t=24$ (lower right panel) with $N_{\rm smr}=70$ as a typical case. The dotted lines with shaded bands indicate the $T=0$ result 
obtained from the simulation with $N_t=80$.
For the case of $T<T_C$ such as
$N_t=72$ (upper left panel)
and $N_t=48$ (upper right panel), 
the standard effective masses (denoted by circle symbols) corresponding to the values of ${M}_{\rm eff}$ are consistent with
the diamond symbols corresponding to $\widetilde{M}_{\rm eff}$. 
Both effective masses are close to the dotted lines.
On the other hand, for the case of $T>T_C$ such as $N_t=32$ (lower left panel) and $N_t=24$ (lower right panel), the standard effective masses are observed to shift below the dotted lines for large $t$.
This trend is consistent with previous studies~\cite{{Ishii:2001zq},{Ishii:2002ww}}, but differs from the effective mass given by the time-differential correlator, which appears well above the dotted line.
This implies that the original two-point function $C(t)$ contains the constant term. 

To confirm the presence of the constant term in $C(t)$, we next employ a correlated fit on
$\widetilde{C}(t)=C(t)/C(0)$
using the following functional form
\begin{align}
\widetilde{C}(t)=C_0+A\cosh[M_G(t-N_t/2)]
\label{Eq:FitForm}
\end{align}
with three parameters, $C_0$, $A$ and $M_G$. The fit results are tabulated in Tables~\ref{tab:fit_mass_T1} and
\ref{tab:fit_mass_T2}.

%
%
\begin{table}[t]
\centering
\caption{
Number of gauge configurations ($N_{\rm cfg}$) are used in 
our lattice simulations at target temperature $T$ with temporal lattice size of $N_t$.
The critical temperature was estimated as $T_c\simeq 280$ MeV in Refs.~\cite{{Ishii:2001zq},{Ishii:2002ww}}.
\label{tab:finite_T_configs}
}
\begin{ruledtabular} 
\begin{tabular}{l l c }
\hline
$N_s^3 \times N_t$ & $T$ [MeV] & $N_{\rm cfg}$\cr
\hline
$20^3 \times 72$ ($24^3 \times 72$, $16^3 \times 72$)& 130 $(T\approx 0.5T_c)$ & 2000 \cr 
$20^3 \times 48$ & 195 $(T\approx 0.75T_c)$ & 4000 \cr 
$20^3 \times 40$ & 234  & 4000 \cr 
$20^3 \times 36$ & 260  & 10000 \cr 
$20^3 \times 34$ & 275 $(T\approx T_c)$ & 7500 \cr 
$20^3 \times 32$ & 293  & 7500 \cr 
$20^3 \times 28$ & 334 & 7500 \cr 
$20^3 \times 24$ ($24^3 \times 24$, $16^3 \times 24$) & 390 $(T\approx 1.5T_c)$& 7500 \cr 
$20^3 \times 20$ & 468 & 7500 \cr 
\hline
\end{tabular}
\end{ruledtabular}
\end{table}

\begin{figure}[t]
\includegraphics[width=0.48\linewidth,bb=0 0 792 612,clip]{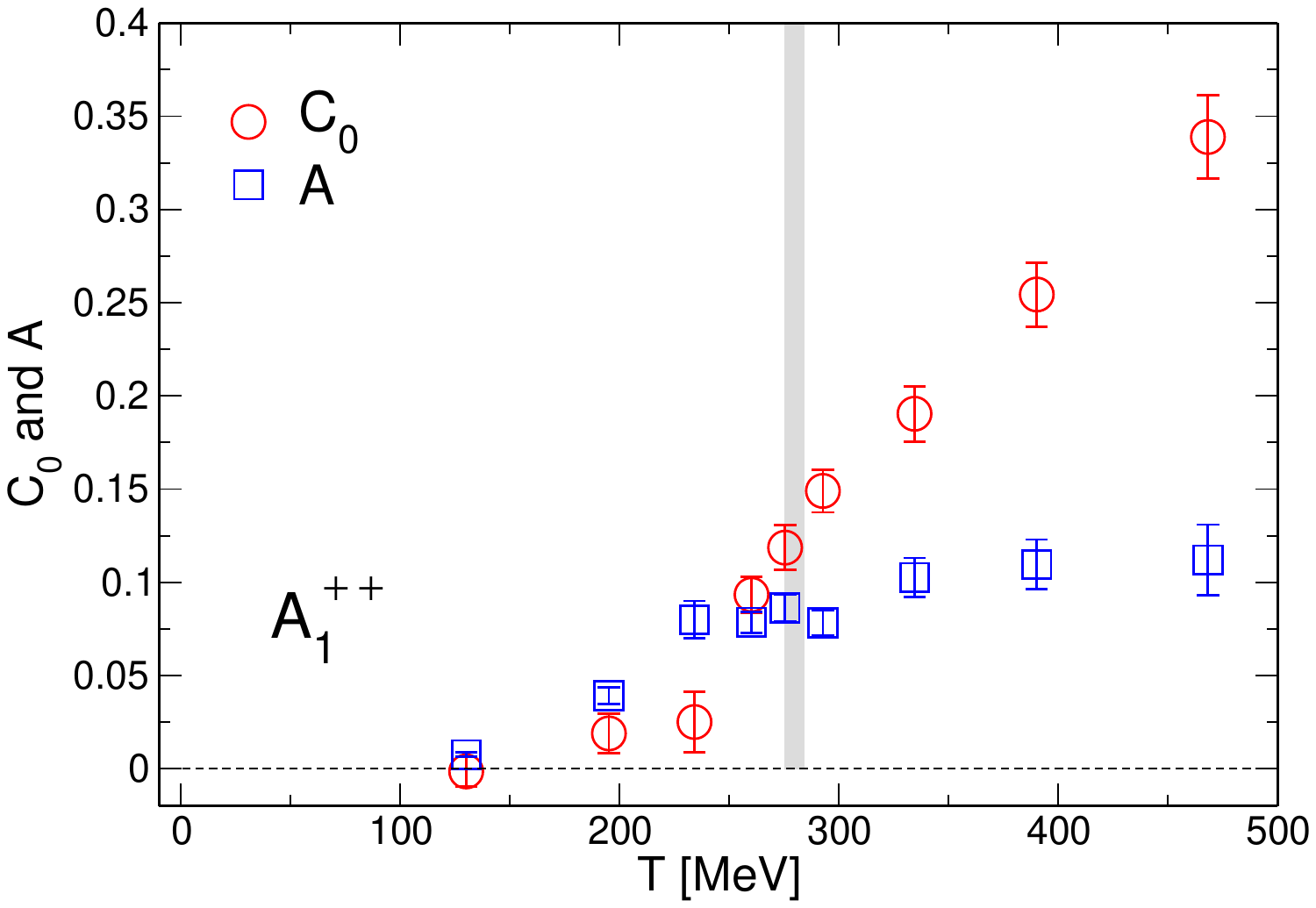}
\includegraphics[width=0.48\linewidth,bb=0 0 792 612,clip]{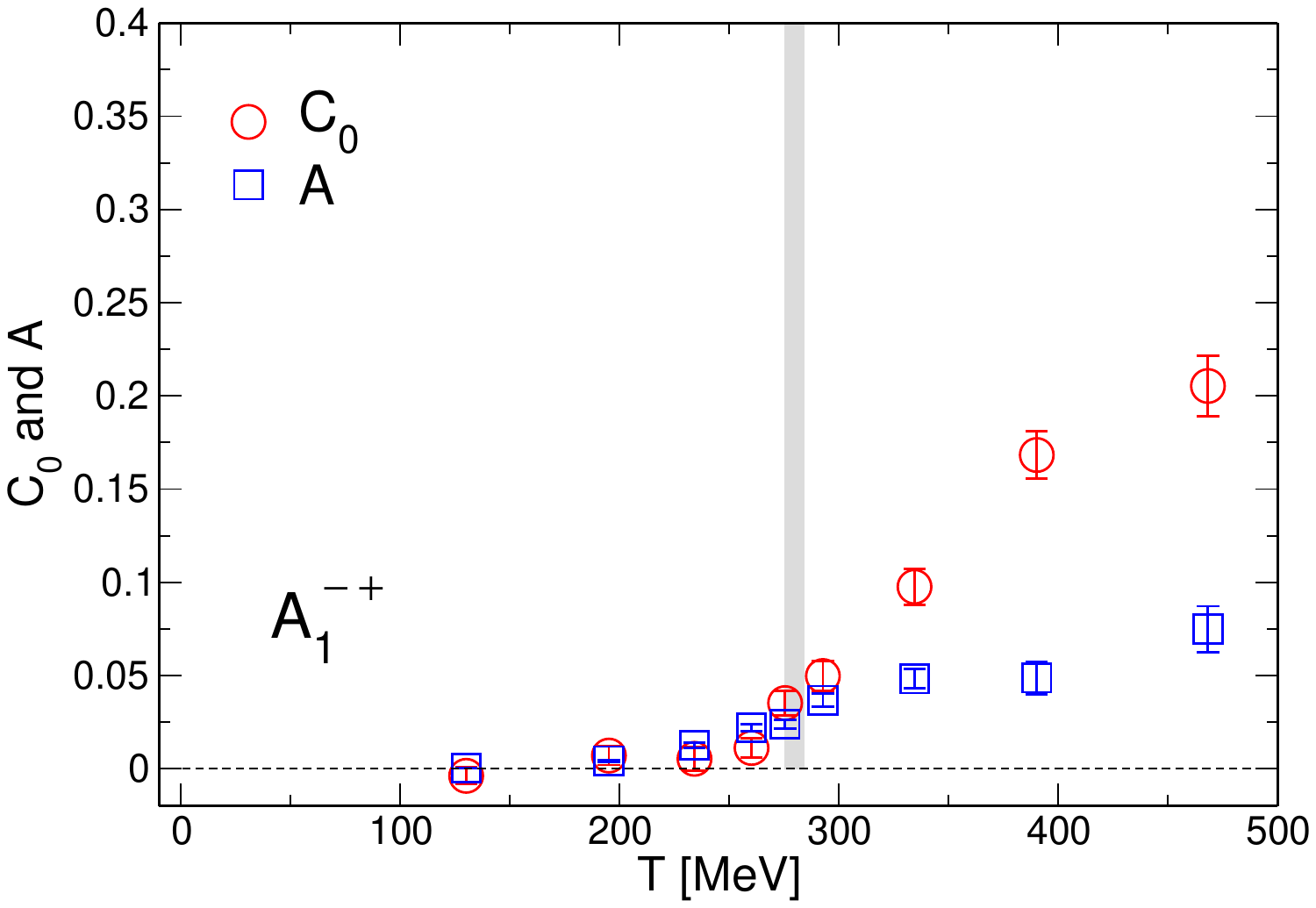}
\includegraphics[width=0.48\linewidth,bb=0 0 792 612,clip]{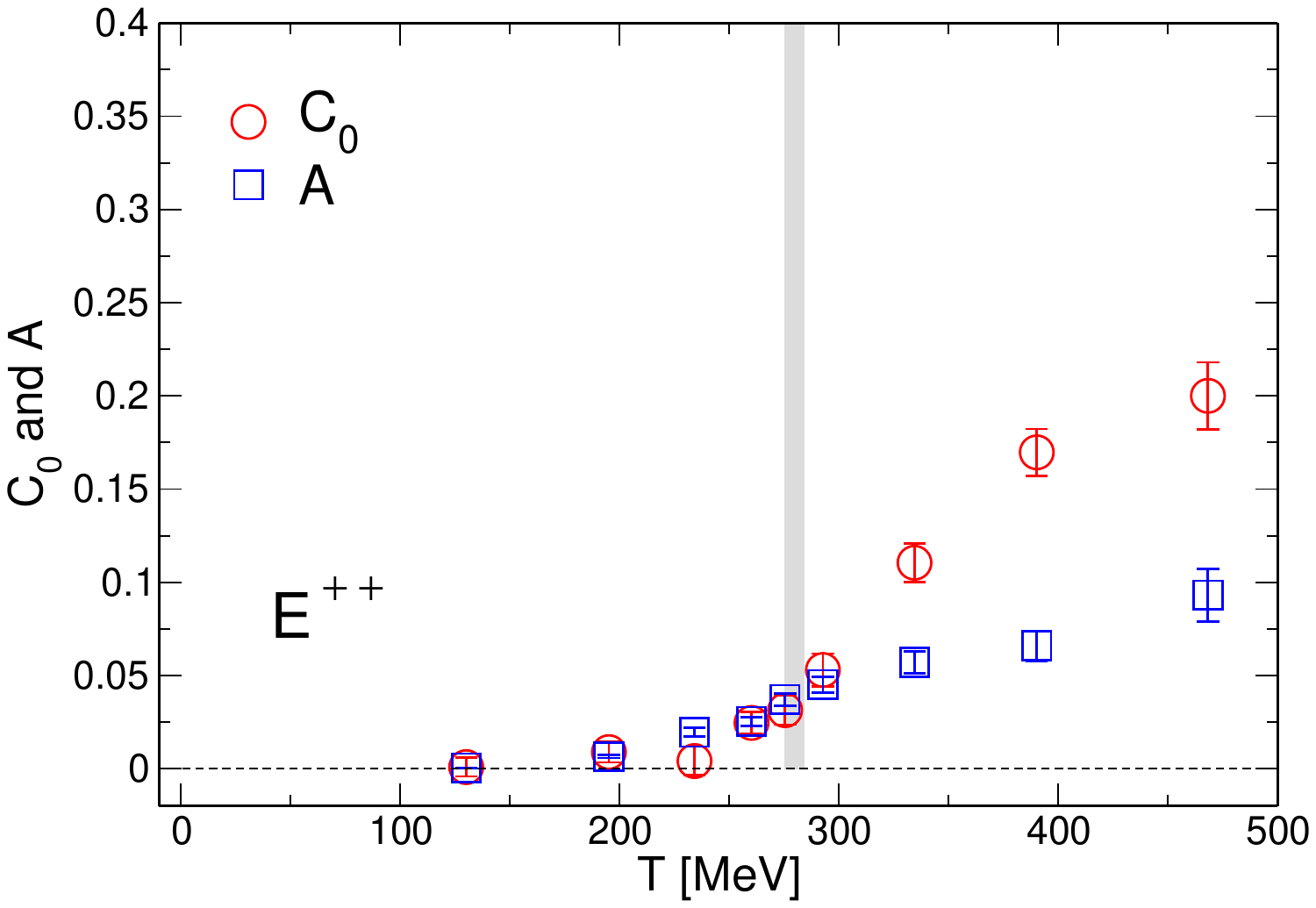}
\includegraphics[width=0.48\linewidth,bb=0 0 792 612,clip]{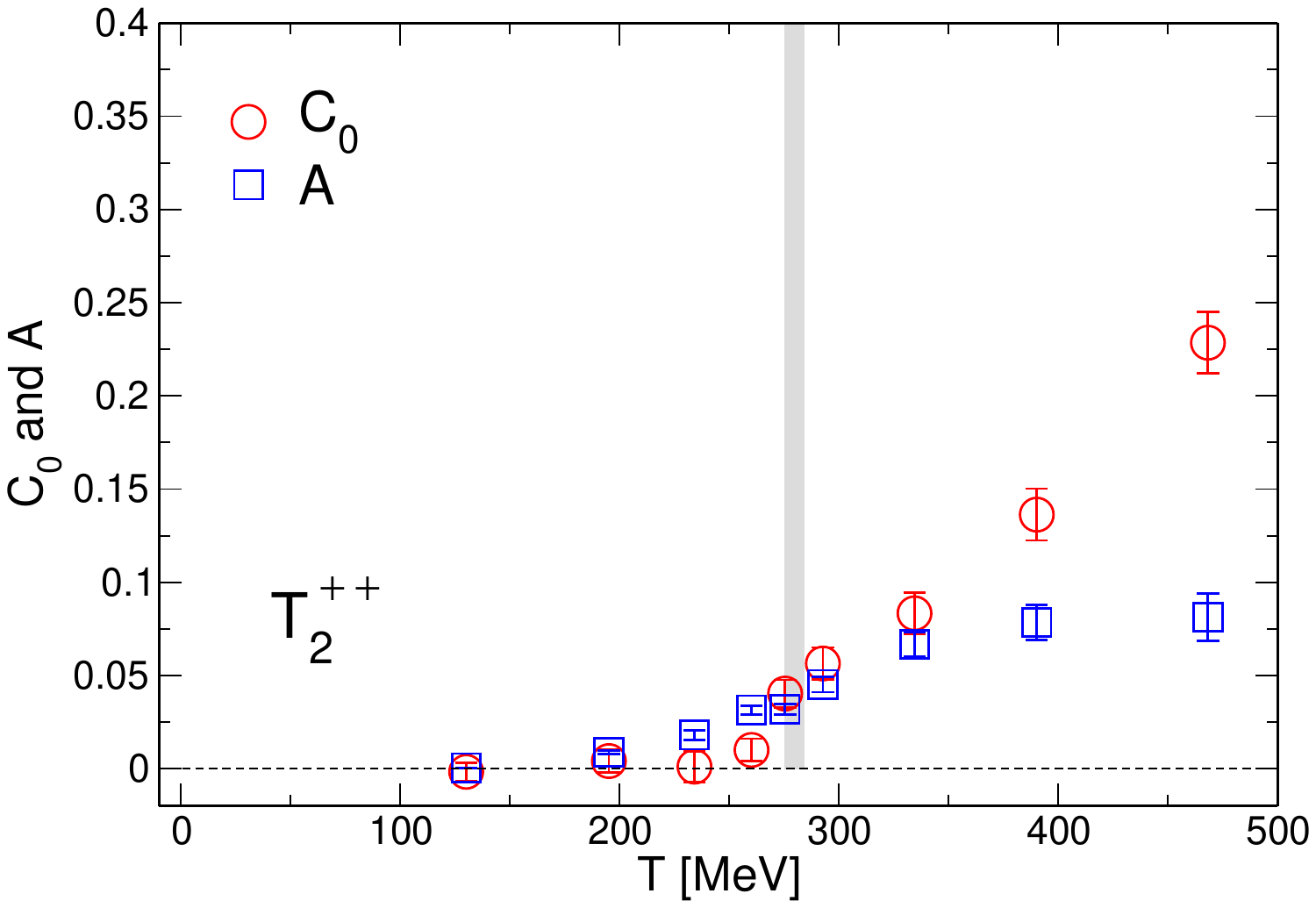}
 \caption{
 Fit results of $C_0$ (red circles) and $A$ (blue squares)
 as a function of temperature $T$
 for the $A_1^{++}$ (upper left panel), $A_1^{-+}$ (upper right panel), $E^{++}$ (lower left panel) and $T_2^{++}$ (lower right panel) irreps. The vertical gray-shaded band denotes the critical temperature, $T_C\approx 280$ MeV.
\label{fig:C_and_A_temp}}
\end{figure}

In Fig.~\ref{fig:C_and_A_temp},
we plot the fit results of $C_0$ compared with the value of $A$ as a function of $T$. A vertical gray-shaded band denotes the 
critical temperature, $T_C\approx 280$ MeV. In the range of $T<T_C$, the value of $C_0$ is observed to be consistent with zero within statistical uncertainty, while in the range above $T_C$ a non-zero value of $C_0$ is identified in all four channels. Indeed, the fit results strongly suggest that there is a constant term in $C(t)$ at temperatures above $T_C$. 

\begin{figure}[t]
\includegraphics[width=0.48\linewidth,bb=0 0 792 612,clip]{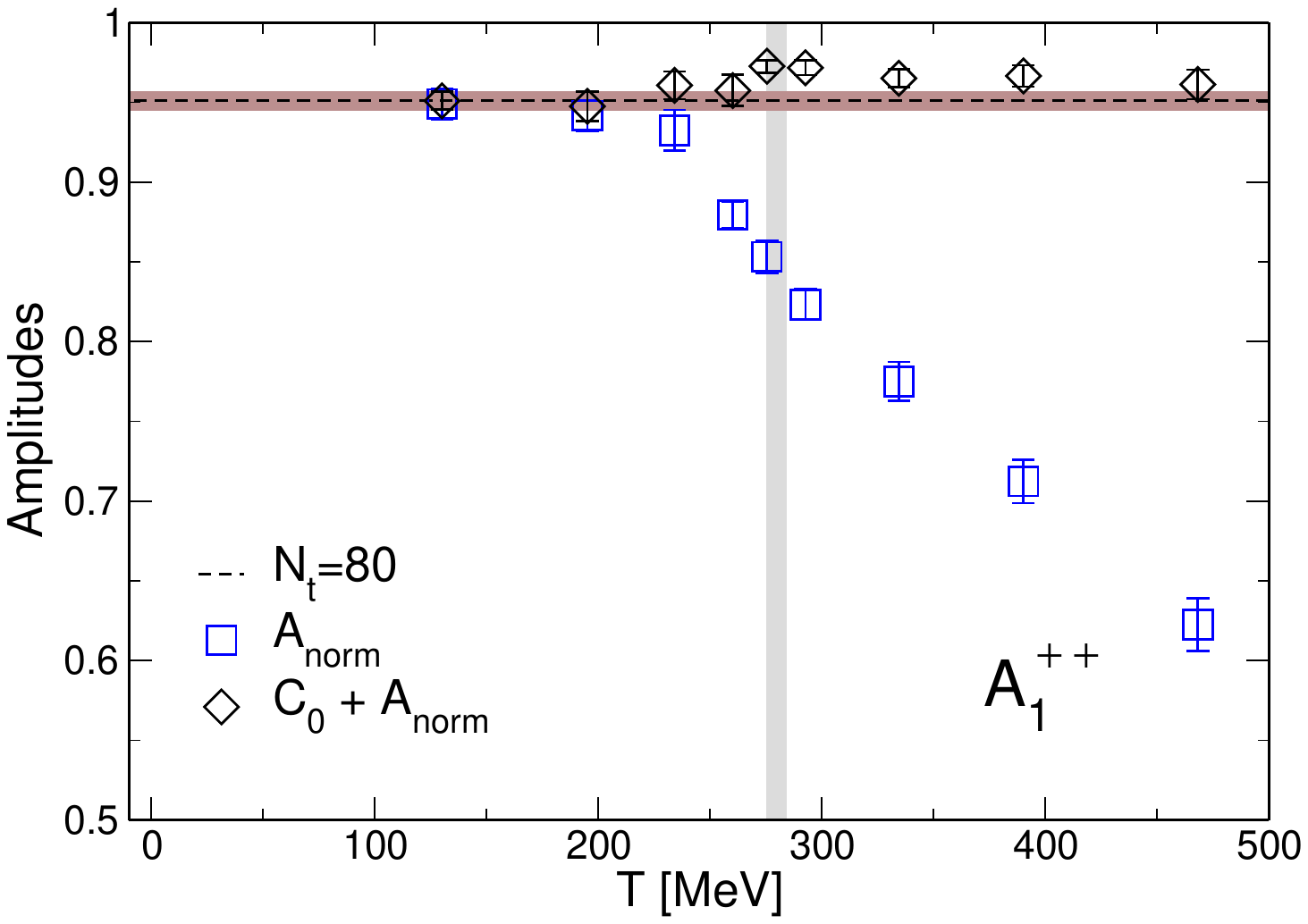}
\includegraphics[width=0.48\linewidth,bb=0 0 792 612,clip]{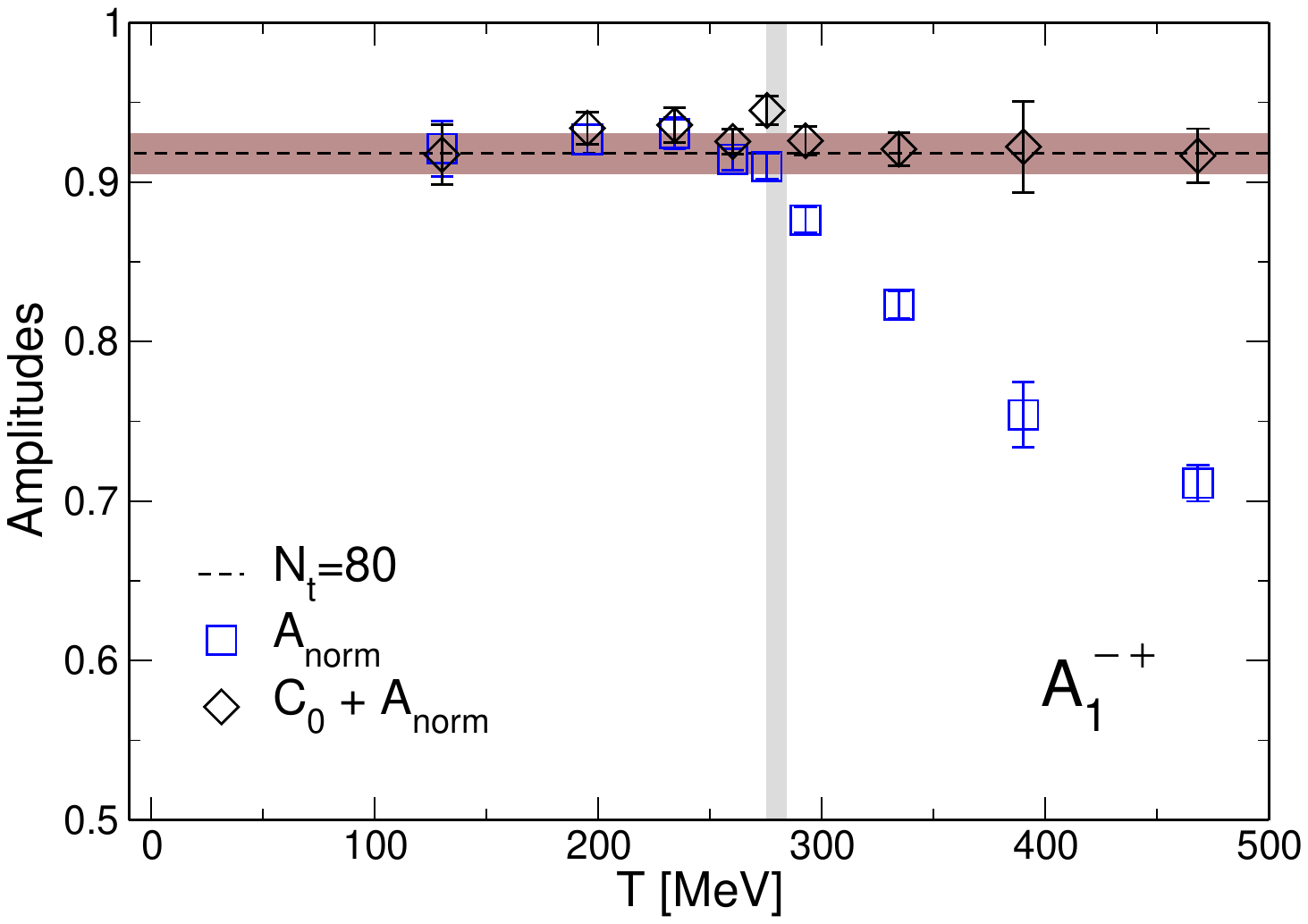}
\includegraphics[width=0.48\linewidth,bb=0 0 792 612,clip]{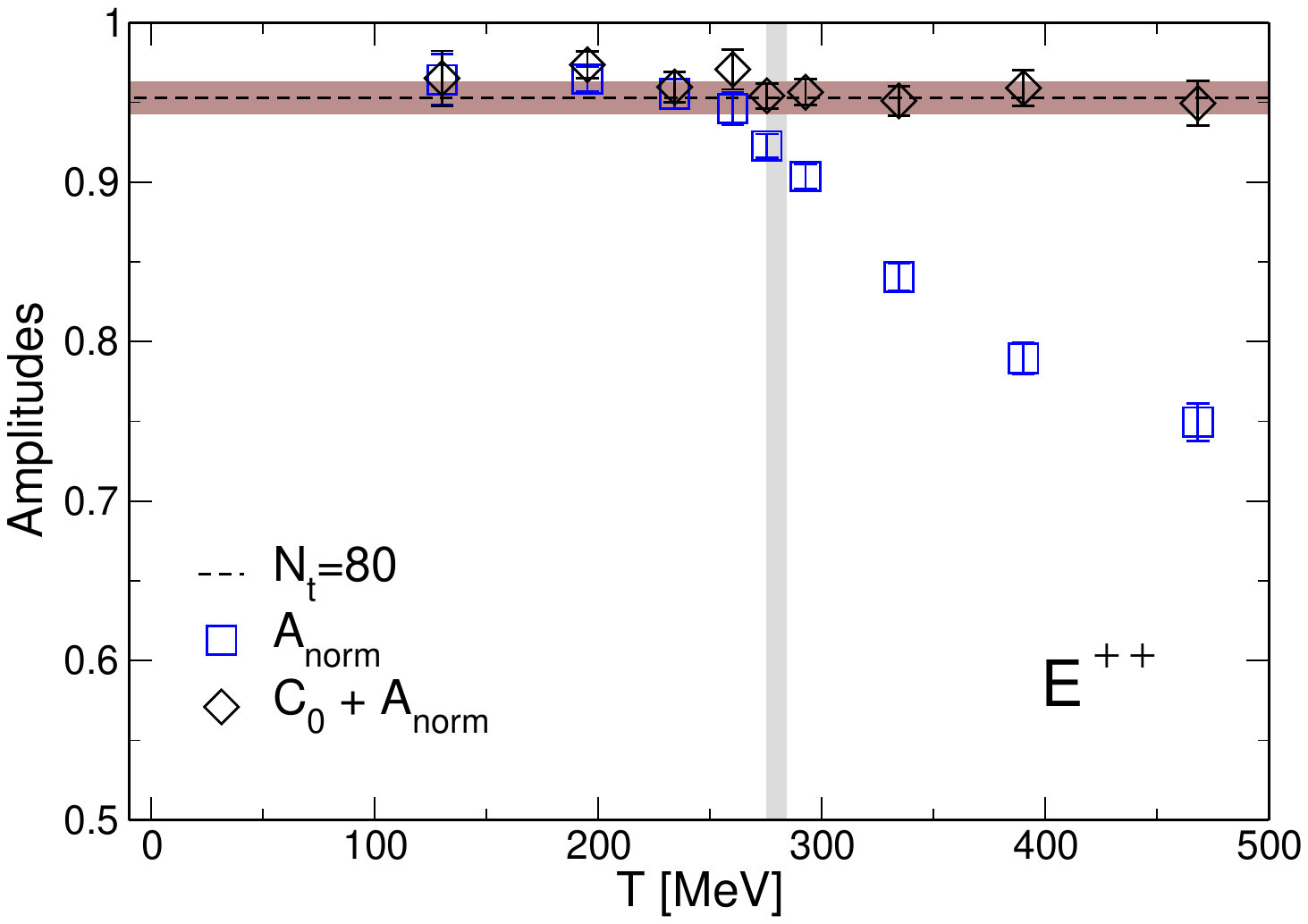}
\includegraphics[width=0.48\linewidth,bb=0 0 792 612,clip]{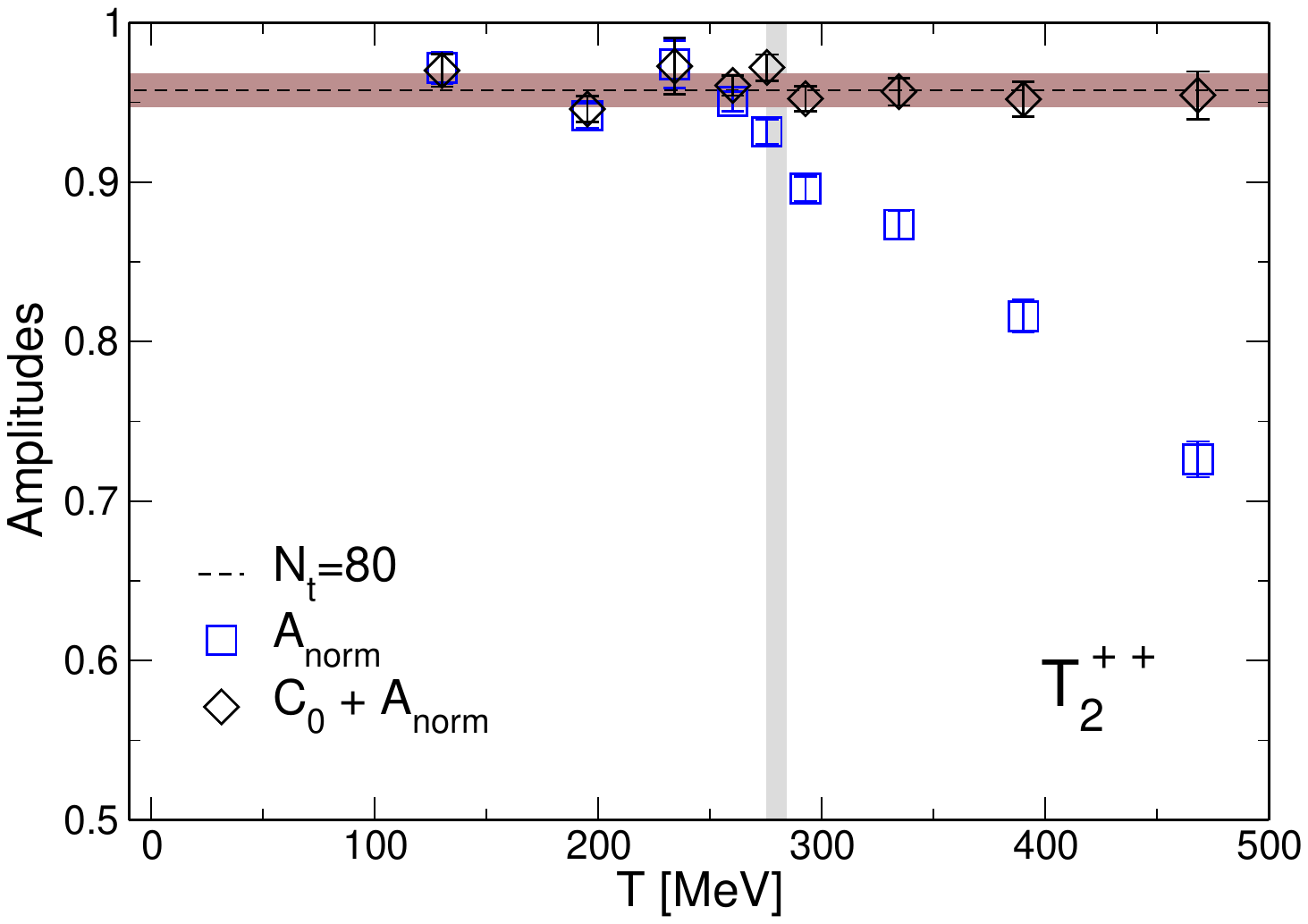}
 \caption{
Normalized amplitude $A_{\rm norm}$ and
total sum of $C_0+A_{\rm norm}$ for 
the $A_1^{++}$ (upper left panel), $A_1^{-+}$ (upper right panel), $E^{++}$ (lower left panel) and $T_2^{++}$ (lower right panel) irreps as a function of temperature $T$. The dotted horizontal lines with brown-shaded bands indicate the corresponding normalized amplitudes obtained for $N_t=80$ as $T=0$. 
The vertical gray-shaded bands denote the critical temperature, $T_C\approx 280$ MeV.
\label{fig:CplusA_norm}
}
\end{figure}

Figure~\ref{fig:CplusA_norm} illustrates how the glueball ground-state contribution varies with temperature in the two-point function due to the manifestation of the constant term.
In all four channels, the normalized amplitude $A_{\mathrm{norm}}$ 
shows no change from zero temperature to near the phase transition temperature, but suddenly begins to monotonically decrease with increasing temperature from near the phase transition point. 
In particular, the lowest-lying scalar ($A_1^{++}$) glueball, shows the strongest tendency compared to the other glueball states. Furthermore, it is observed that the sum of $C_0$ and $A_{\mathrm{norm}}$ shows no change with increasing temperature.
This implies that $C_0$ can be regarded as the ``zero-energy'' contribution that is normalized regardless of the size of $N_t$. Indeed, since it is the ``zero-energy'' contribution, the variational analysis~\cite{{Michael:1985ne},{Luscher:1990ck}} using
the correlation matrix constructed by the $\widetilde{\cal O}_{\rm GB}$ operator with different
smearing steps could not separate the constant contribution
from the finite energy modes.

%
%
%

%
%

\subsection{Glueball mass at finite temperature}
\begin{figure}[t]
\includegraphics[width=0.48\linewidth,bb=0 0 792 612,clip]{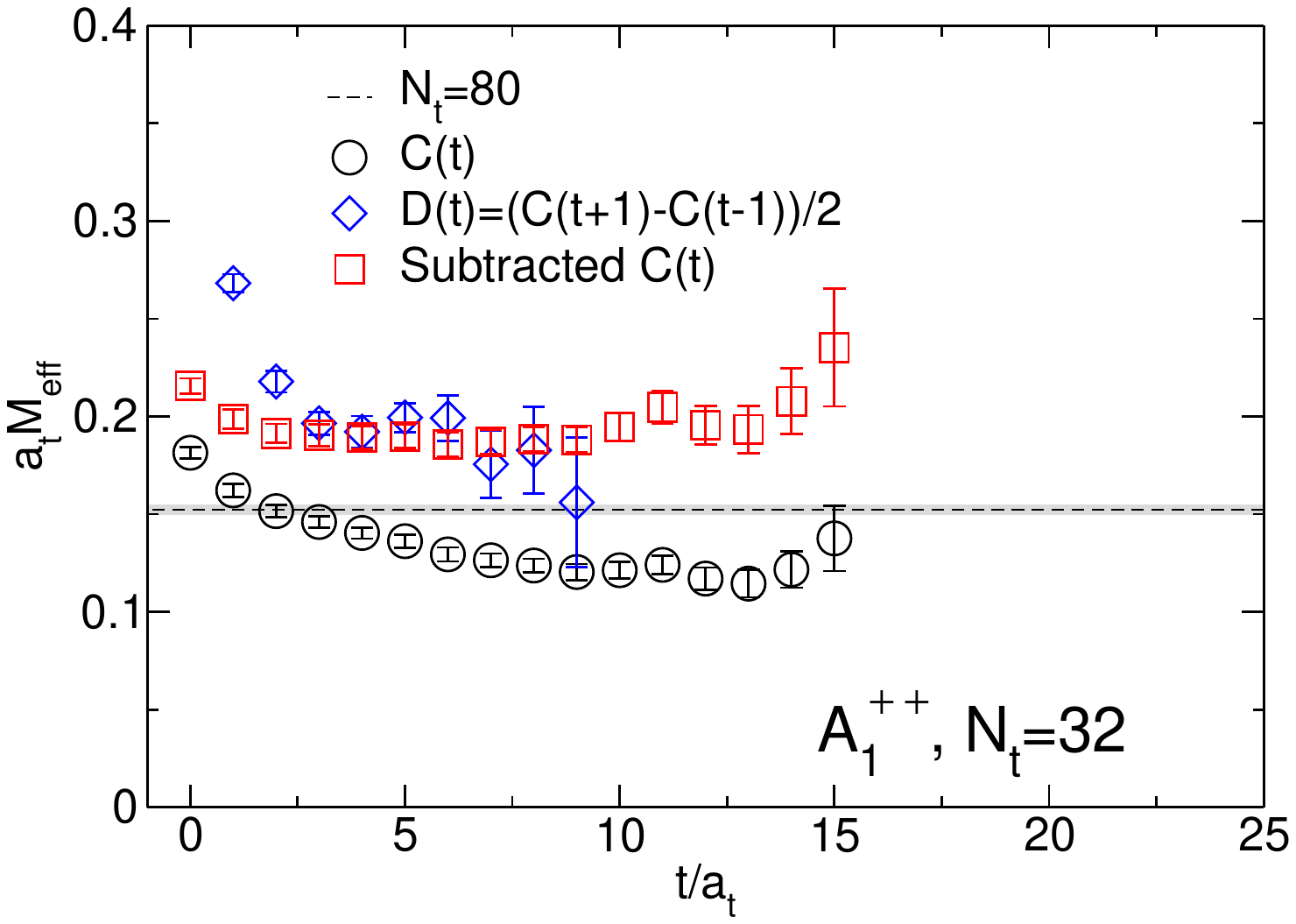}
\includegraphics[width=0.48\linewidth,bb=0 0 792 612,clip]{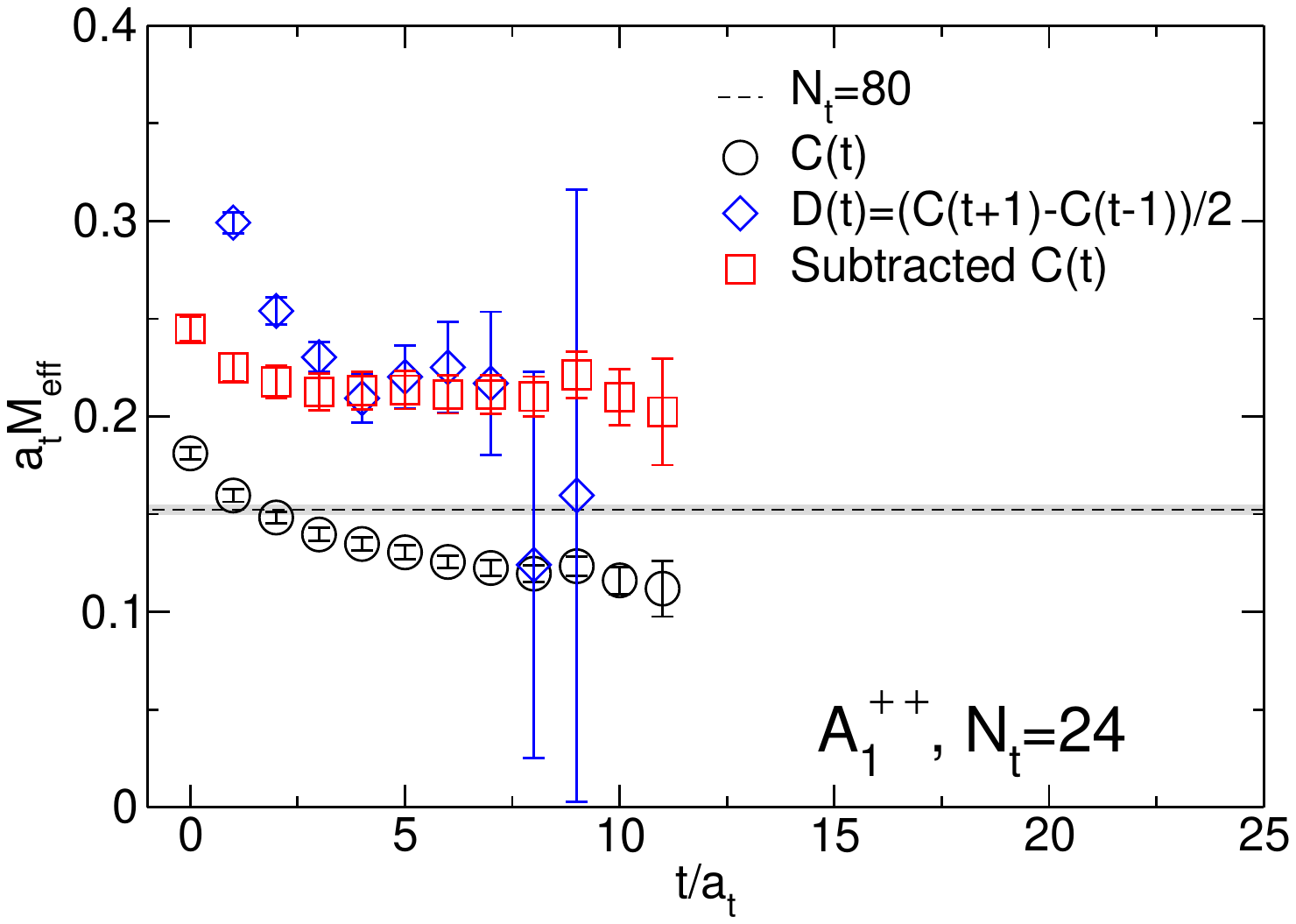}
 \caption{
Effective mass plots for the $A_1^{++}$ irrep for $T>T_C$. The left panel shows the case for $N_t=32$ ($T\approx 1.1T_C$), the right panel shows the case for $N_t=24$ ($T\approx 1.5T_C$).
Red square symbols are given by
the two-point function with the constant term subtracted, while black circle and blue diamond symbols are given by the original
two-point correlator and its time derivative one. Dotted lines with shaded bands indicate the fit result for $N_t=80$ as $T=0$.
\label{fig:Mg_comp_w_subtract_A1pp}}
\end{figure}

\begin{figure}[t]
\includegraphics[width=0.48\linewidth,bb=0 0 792 612,clip]{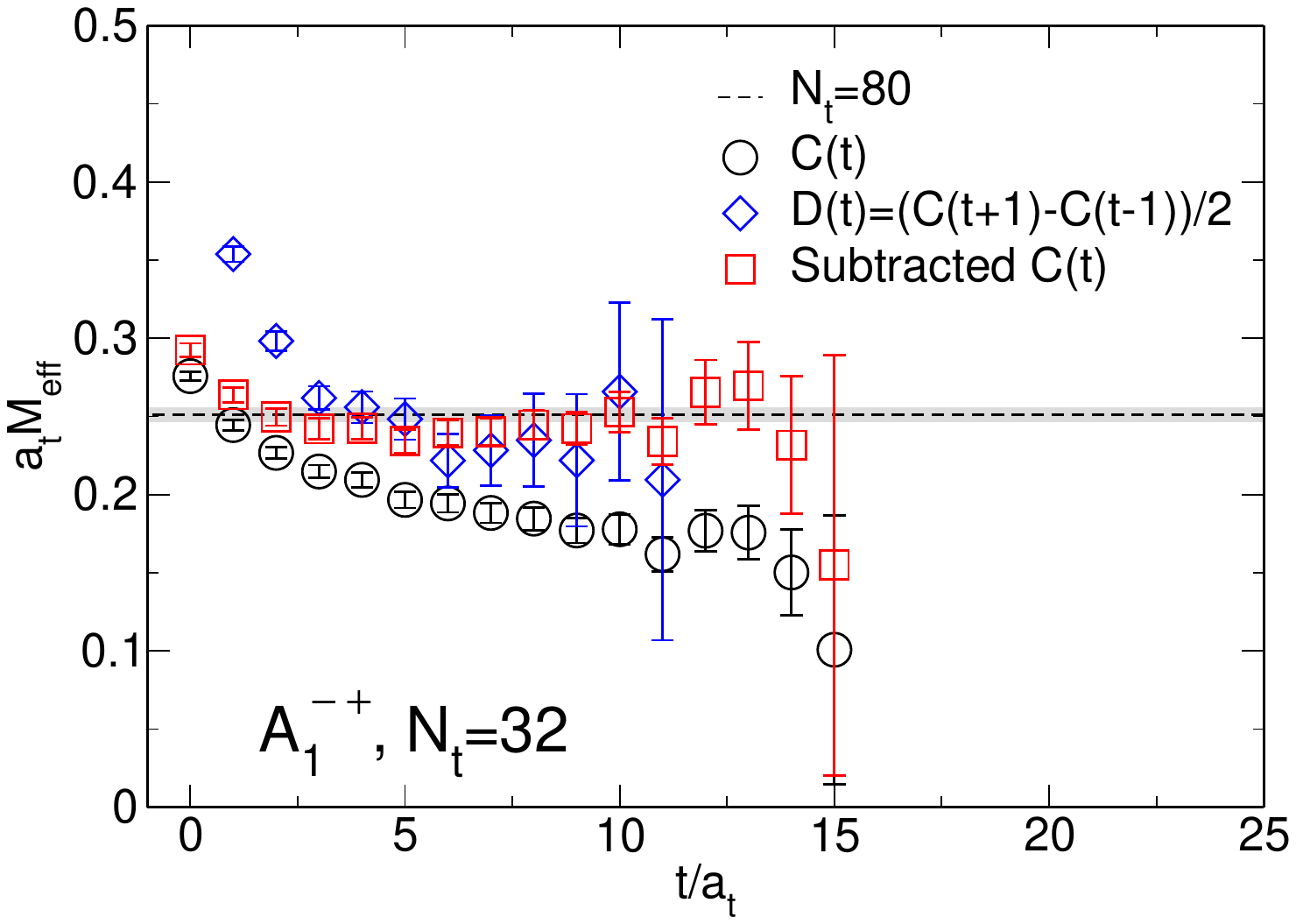}
\includegraphics[width=0.48\linewidth,bb=0 0 792 612,clip]{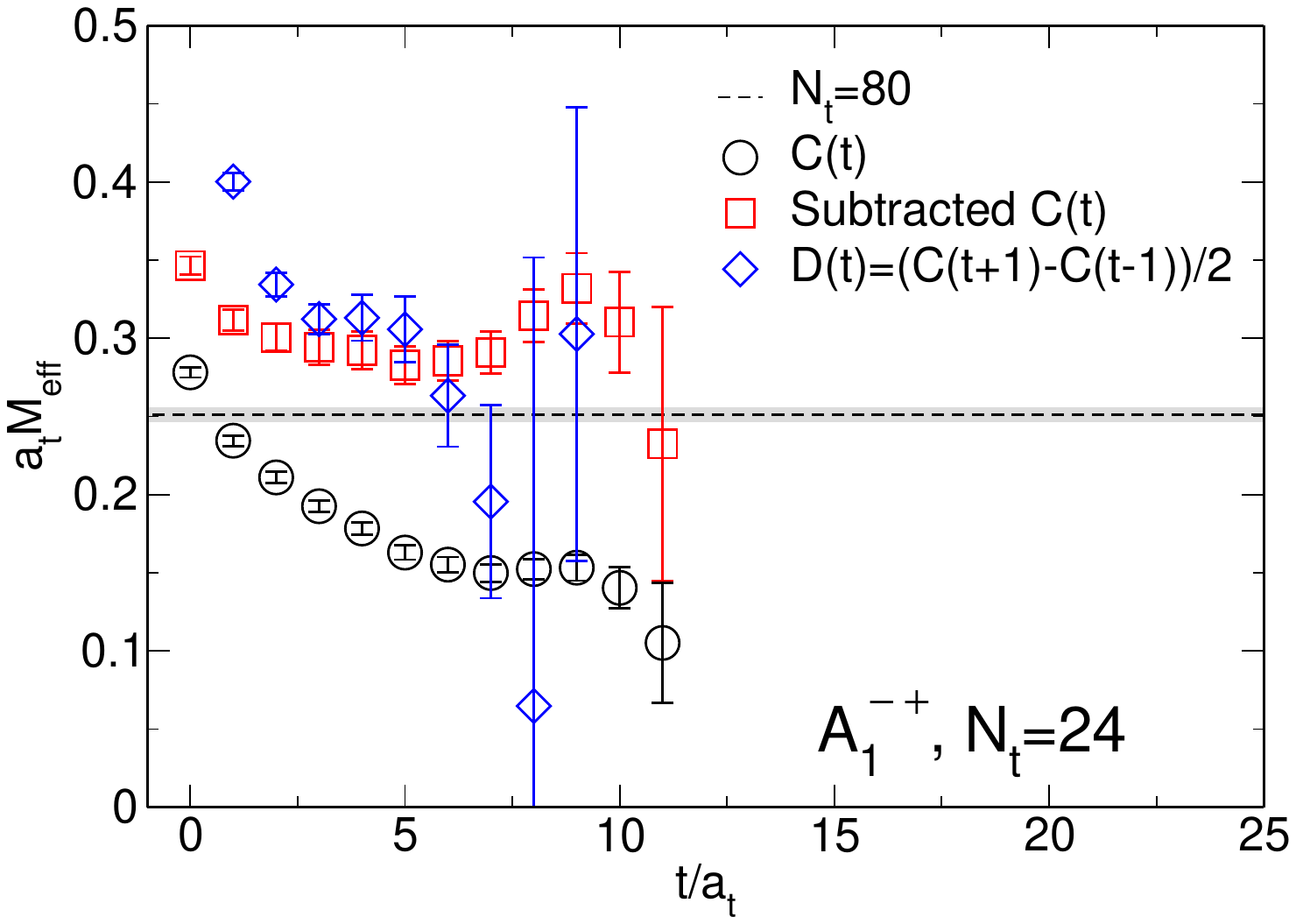}
 \caption{
 Same as Fig.~\ref{fig:Mg_comp_w_subtract_A1pp} for the $A_1^{-+}$ irrep. 
\label{fig:Mg_comp_w_subtract_A1mp}}
\end{figure}

\begin{figure}[h]
\includegraphics[width=0.48\linewidth,bb=0 0 792 612,clip]{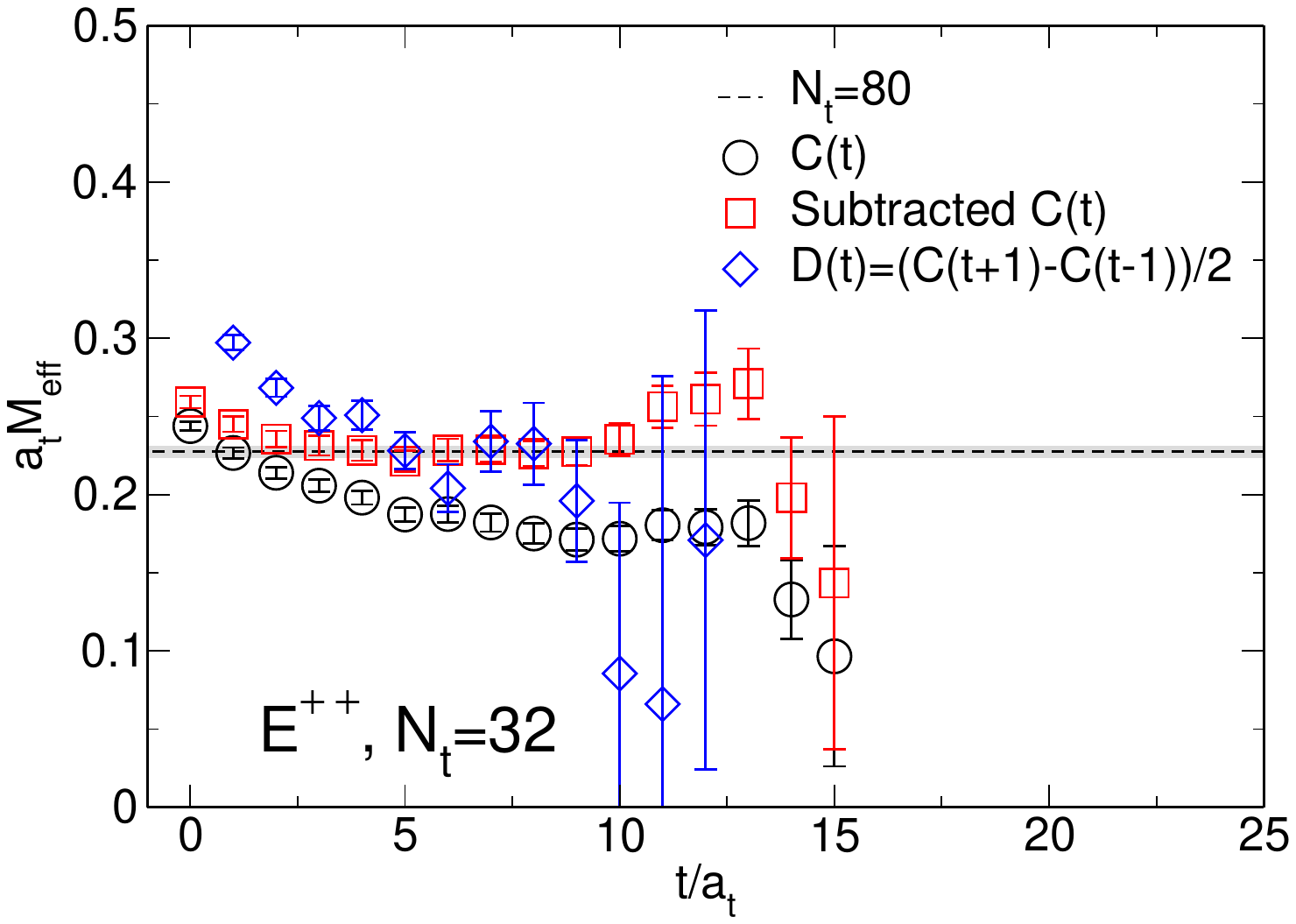}
\includegraphics[width=0.48\linewidth,bb=0 0 792 612,clip]{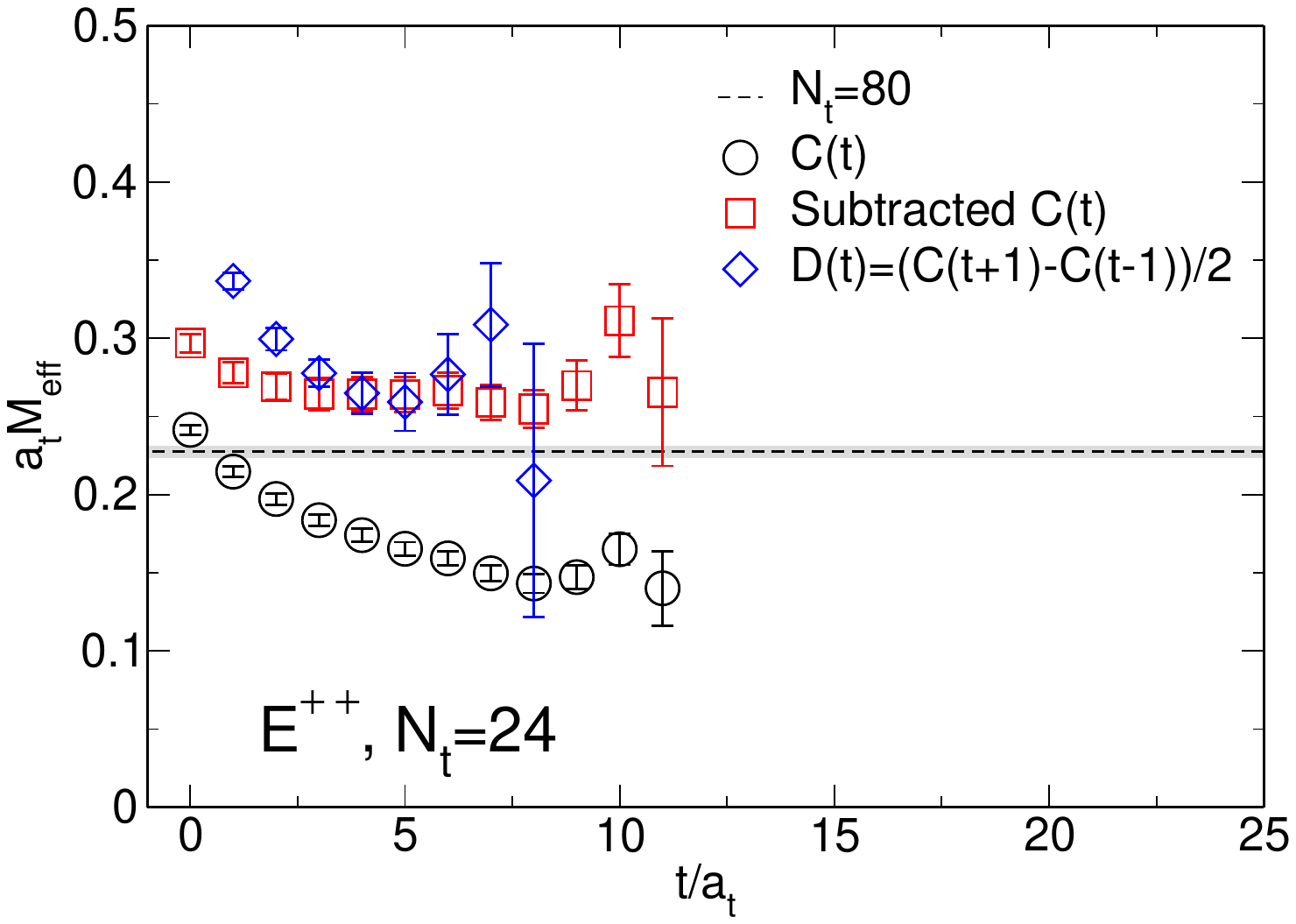}
 \caption{
Same as Fig.~\ref{fig:Mg_comp_w_subtract_A1pp} for the $E^{++}$ irrep. 
\label{fig:Mg_comp_w_subtract_E}}
\end{figure}

\begin{figure}[h]
\includegraphics[width=0.48\linewidth,bb=0 0 792 612,clip]{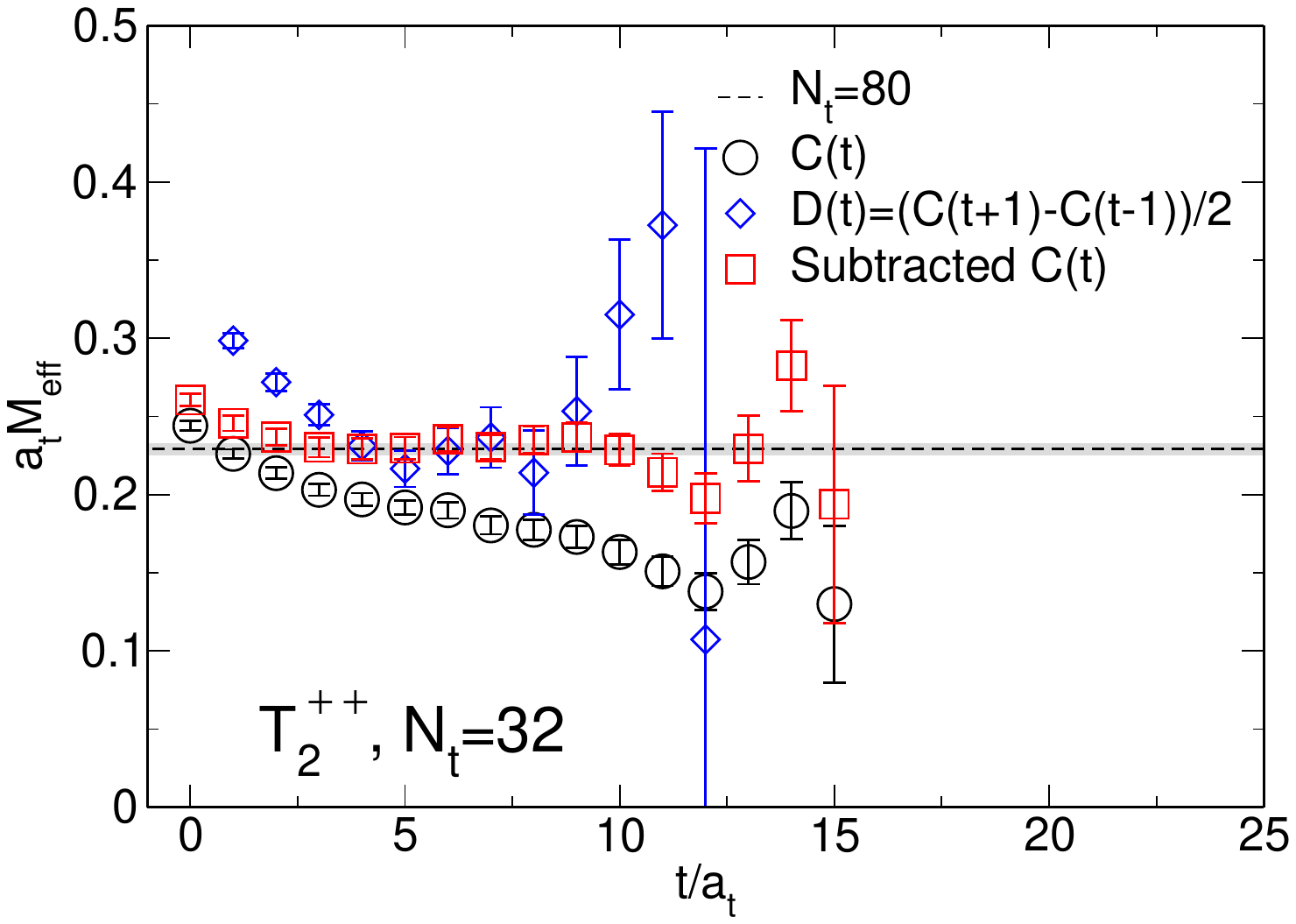}
\includegraphics[width=0.48\linewidth,bb=0 0 792 612,clip]{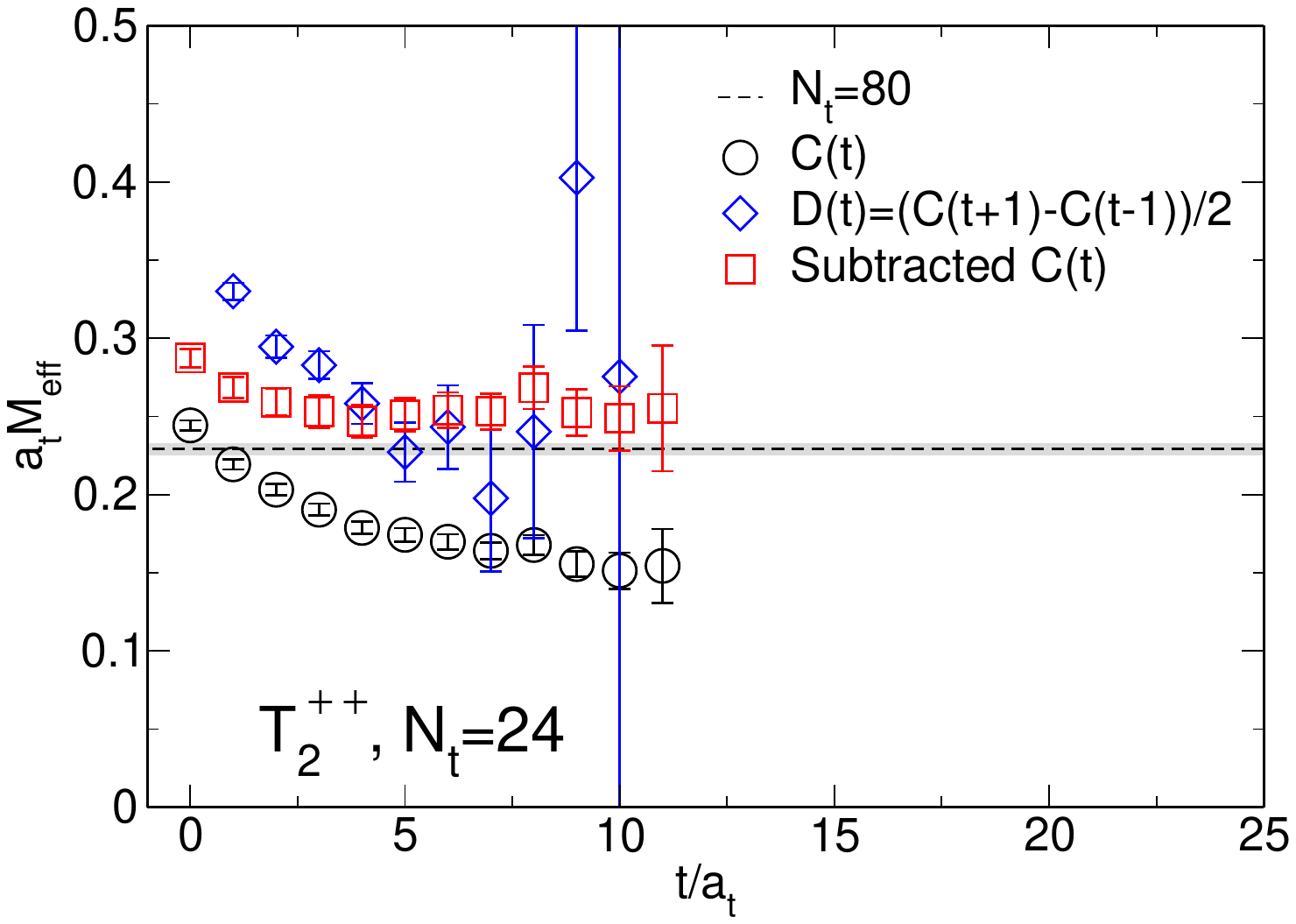}
 \caption{
Same as Fig.~\ref{fig:Mg_comp_w_subtract_A1pp} for the $T_2^{++}$ irrep. 
\label{fig:Mg_comp_w_subtract_T2}}
\end{figure}

Next, let us subtract the constant term from each two-point function using the fit value of $C_0$ and then evaluate the effective mass defined in Eq.~(\ref{eq:effmass_cosh})
for all four channels. The resulting effective mass is plotted over the two results obtained from $C(t)$ and $D(t)$ in each panel of Fig~\ref{fig:Mg_comp_w_subtract_A1pp} through Fig.~\ref{fig:Mg_comp_w_subtract_T2}.
It is evident that the effective mass obtained from the two-point function with the constant term subtracted is in good agreement with the effective mass calculated from $D(t)$ in all four channels. Furthermore, the plateau behavior of the effective mass becomes more pronounced after subtracting the constant term.

\begin{figure}[h]
\includegraphics[width=0.8\linewidth,bb=0 0 792 612,clip]{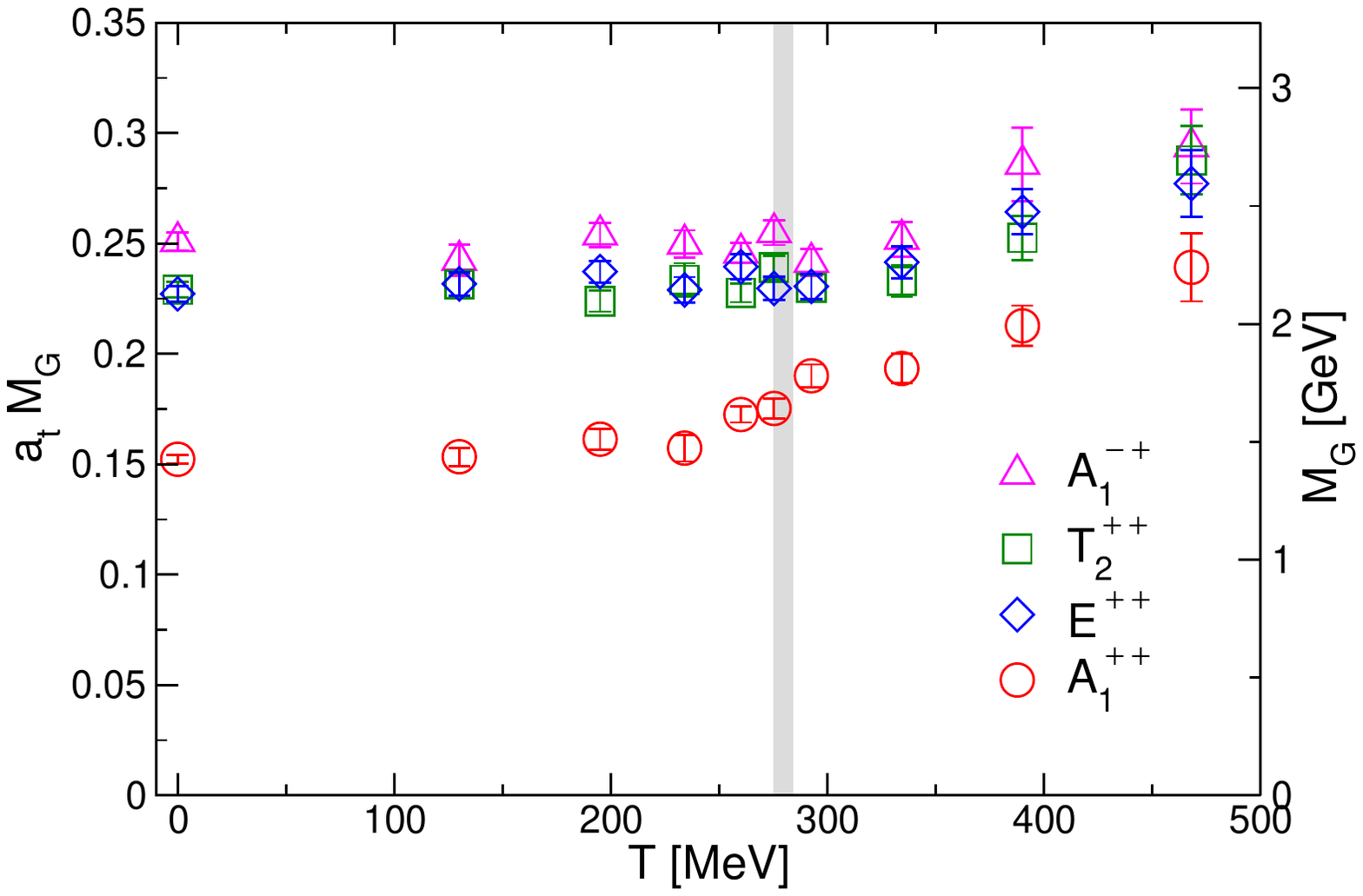}
 \caption{Masses of the glueball ground-states
 for the $A_1^{++}$ (red circles), $E^{++}$ (blue diamonds), $T_2^{++}$ (green squares) and $A_1^{-+}$ (magenta triangles) irreps as a function of temperature $T$. The vertical gray-shaded band denotes the critical temperature, $T_C\approx 280$ MeV.
\label{fig:T_dep_Mass}}
\end{figure}

Finally, the temperature dependence of the glueball ground-state masses for the $A_1^{++}$ (red circles), $E^{++}$ (blue diamonds), $T_2^{++}$ (green squares) and $A_1^{-+}$ (magenta triangles) irreps is shown in Fig.~\ref{fig:T_dep_Mass}.
It is observed that all glueball masses remain almost constant with increasing temperature until just before the phase transition. Subsequent to the phase transition, the mass is also found to increase with increasing temperature $T$. In comparison to tensor glueballs ($E^{++}$ and $T_2^{++}$), and pseudoscalar glueballs ($A_1^{-+}$), the aforementioned trend is particularly pronounced for scalar glueball, whose mass appears to vary linearly with temperature.
Therefore, the mass difference between scalar glueball and the others seems to be smaller in the higher temperature region.

%
%
%

%
%
%

%
%
%

%
%
%

\subsection{Volume dependence of the spectral weight}

The analysis performed in the previous two subsections shows that the constant term appears near the phase transition point, and the mass modification of the glueball states is definitely triggered by the phase transition. Therefore, the inquiry arises as to whether the glueball state  persists as a bound state of gluons in the deconfined phase subsequent to the phase transition. 

If the ``glueball'' becomes a multi-particle state of liberated gluons instead of a single-hadron state, it is anticipated that the volume dependence of the two-point function may exhibit discrepancies between the low- and high-temperature sides across the phase transition. These discrepancies can be attributed to the different normalization conditions applied to single-particle and  multi-particle states in a finite box.
For the single-particle case, 
the two-point function may be scaled by $1/V_s$ with the spatial volume factor $V_s=N_s^3$, while
the two-particle states may  
receive an extra factor of $1/V_s$
in their spectral weigh. See  Appendix~B in Ref.~\cite{Sasaki:2006jn} for more details.

\begin{figure}[b]
\includegraphics[width=0.4\linewidth,bb=0 0 792 612,clip]{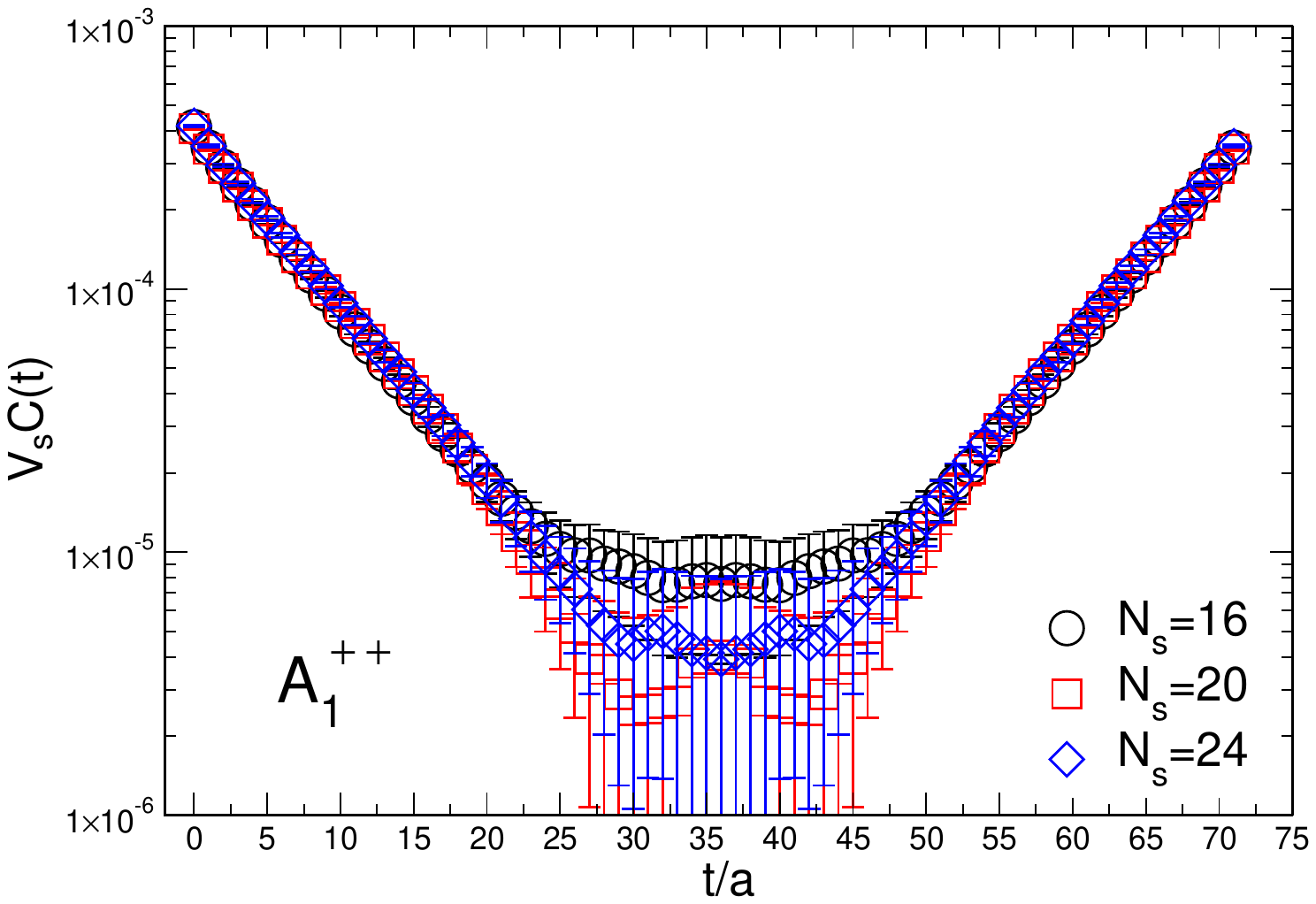}
\includegraphics[width=0.4\linewidth,bb=0 0 792 612,clip]{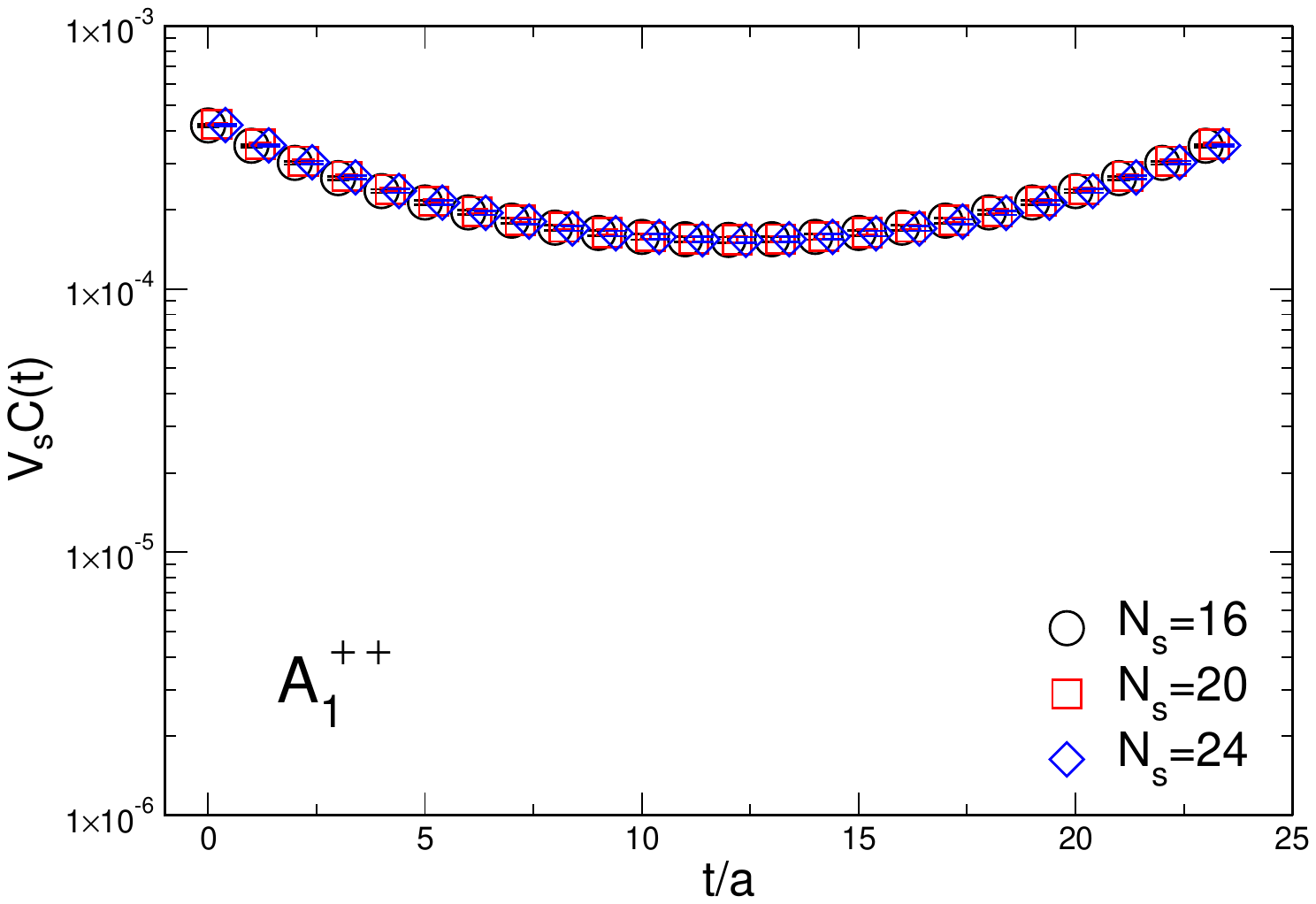}
\includegraphics[width=0.4\linewidth,bb=0 0 792 612,clip]{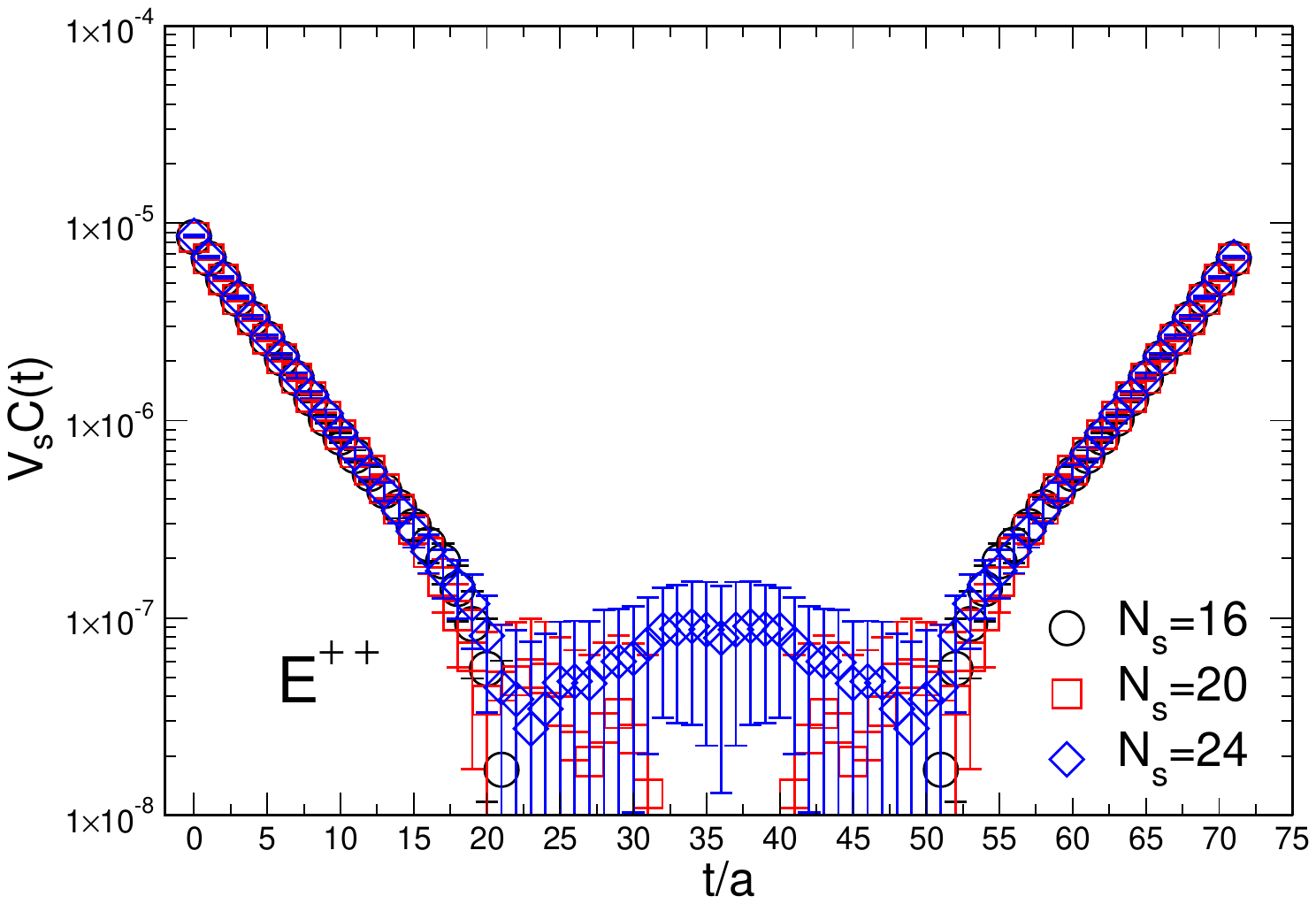}
\includegraphics[width=0.4\linewidth,bb=0 0 792 612,clip]{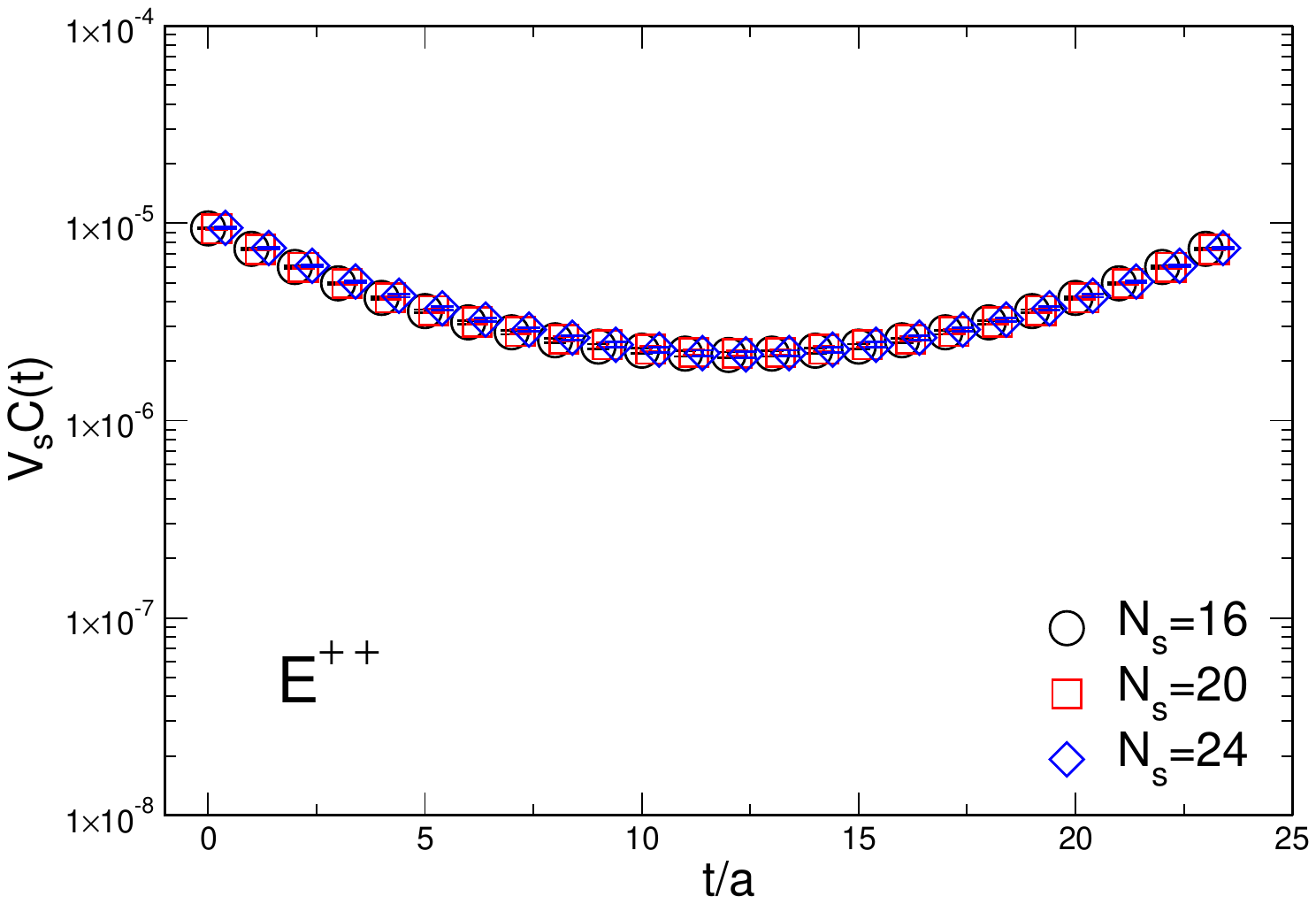}
\includegraphics[width=0.4\linewidth,bb=0 0 792 612,clip]{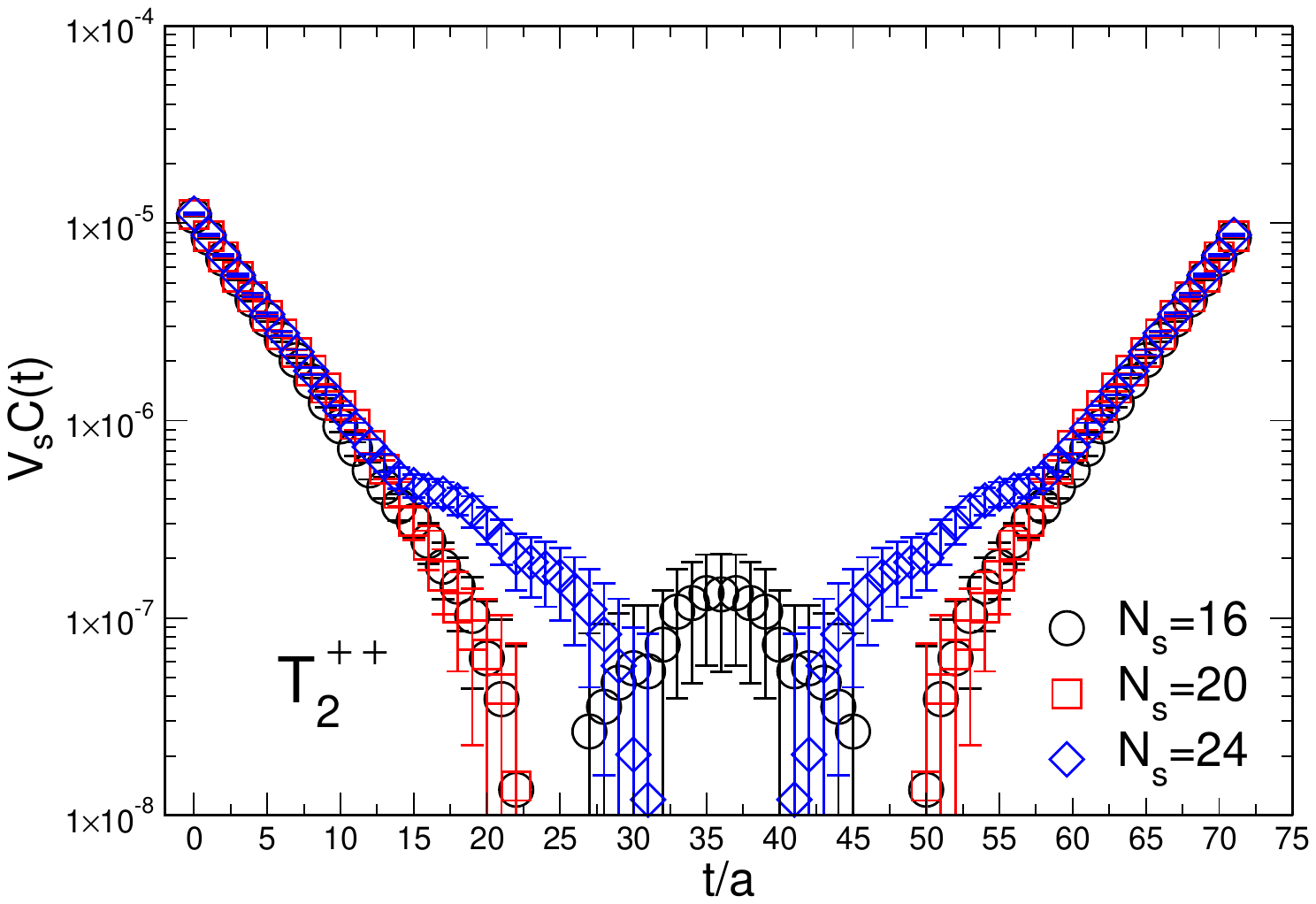}
\includegraphics[width=0.4\linewidth,bb=0 0 792 612,clip]{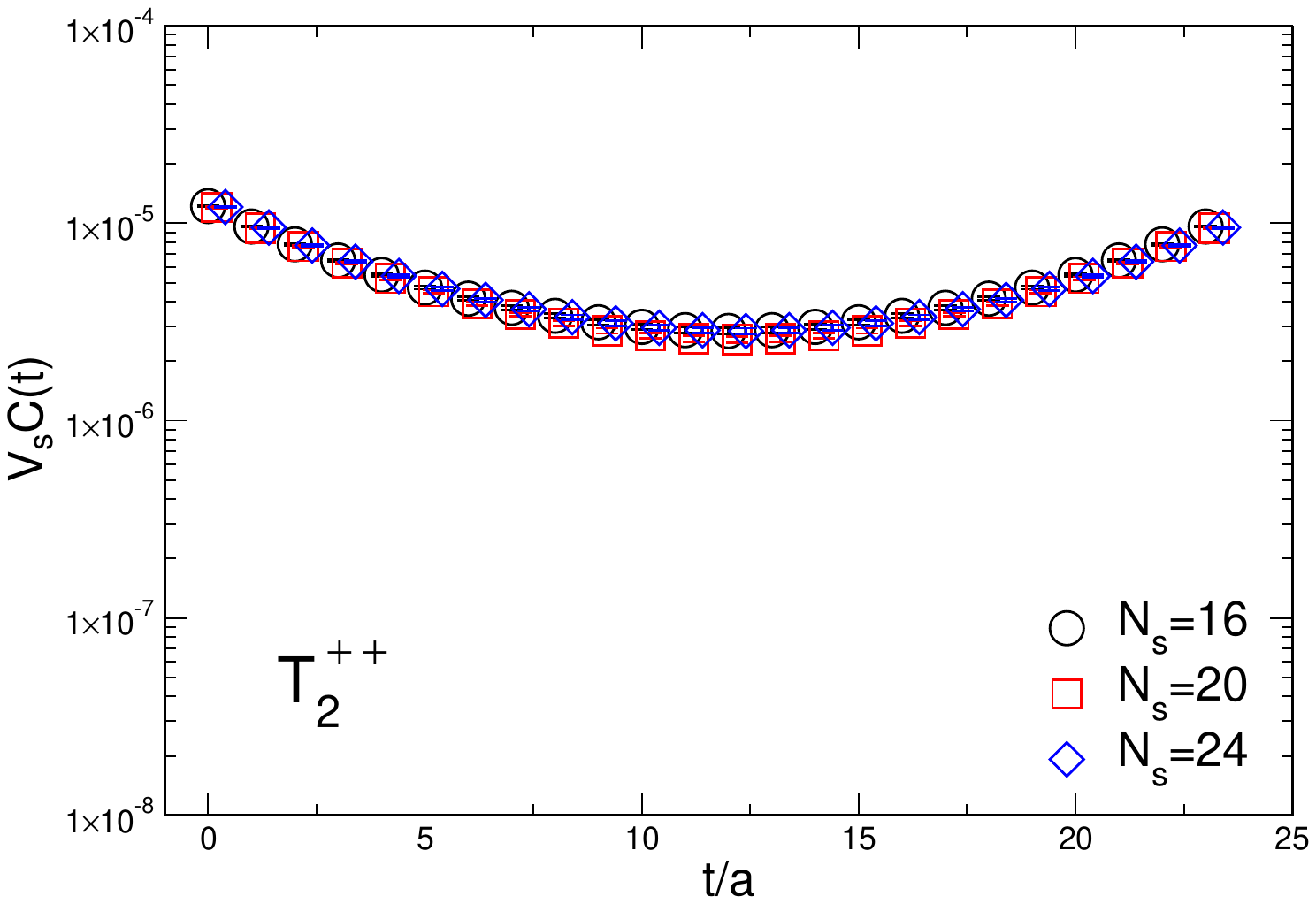}
\includegraphics[width=0.4\linewidth,bb=0 0 792 612,clip]{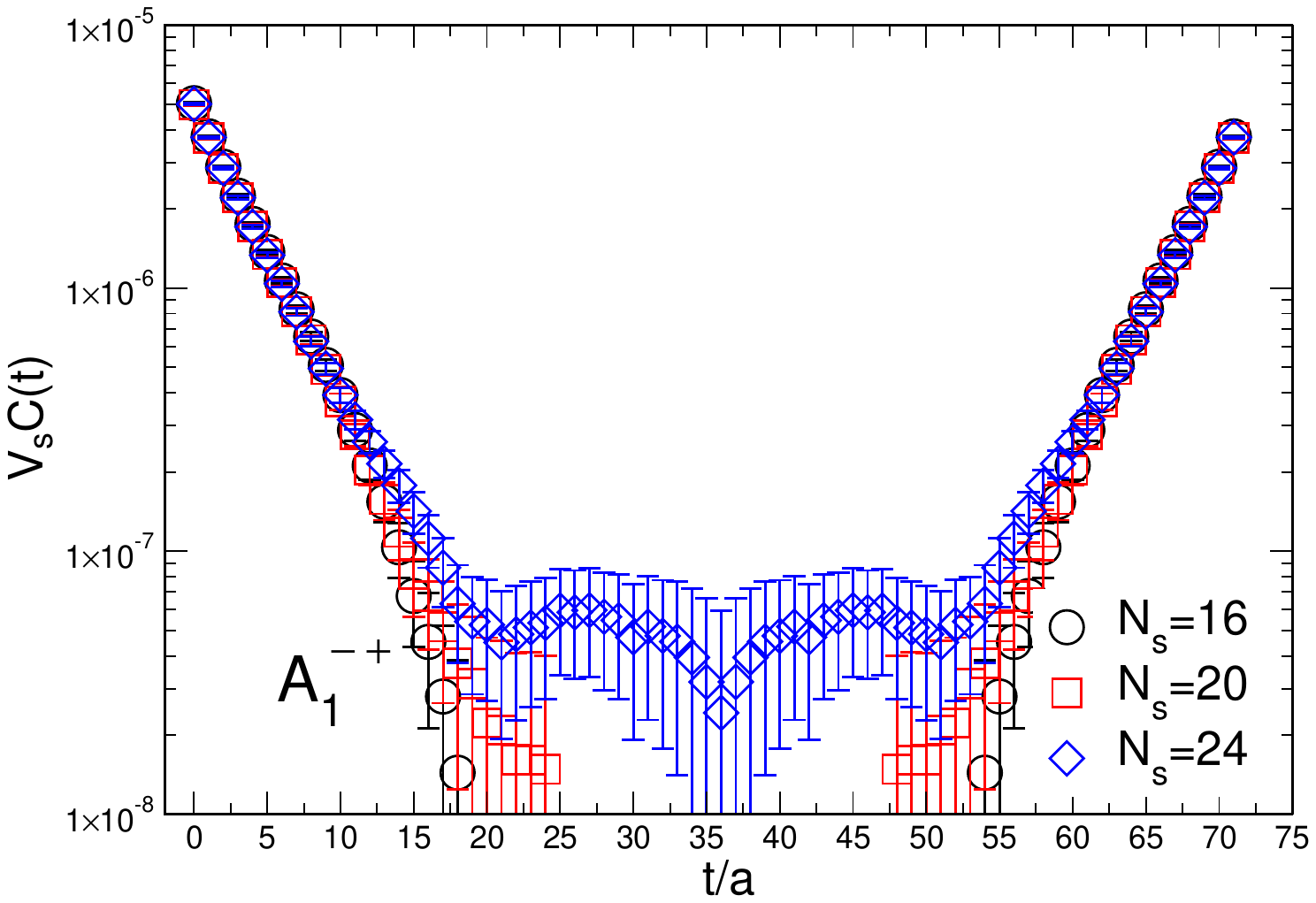}
\includegraphics[width=0.4\linewidth,bb=0 0 792 612,clip]{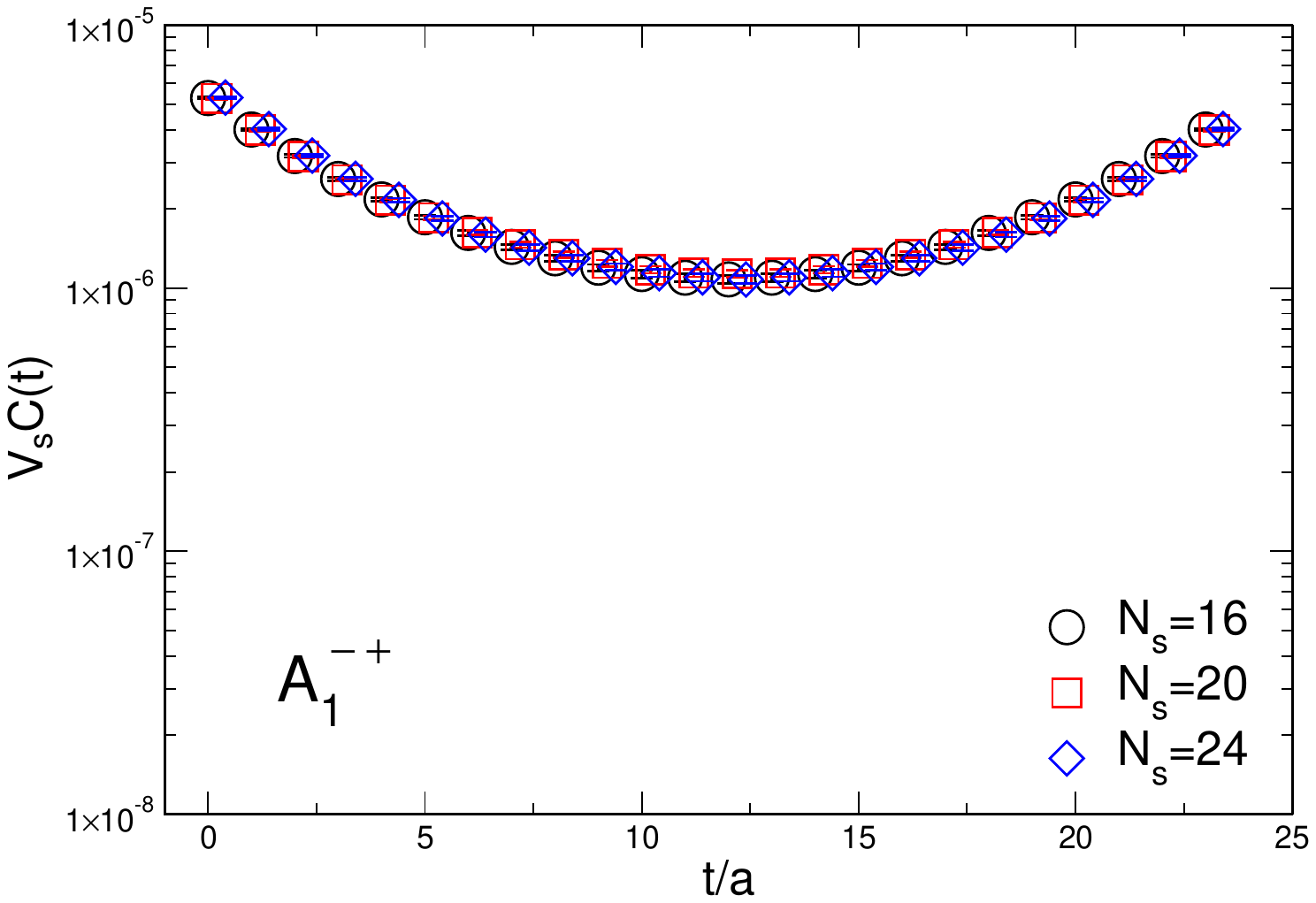}
 \caption{
Finite volume dependence of the two-point function on the low and high temperature sides. The left panel shows the case for $N_t=72$ ($T\approx 0.5T_C$) and the right panel shows the case for $N_t=24$ ($T\approx 1.5T_C$). 
\label{fig:FV_2pt_norm}}
\end{figure}

In Fig.~\ref{fig:FV_2pt_norm}, 
the two-point functions $C(t)$ multiplied by the spatial volume factor $V_s$ are shown on both the low and high temperature sides.
The left panel shows the case for $N_t=72$ ($T\approx 0.5T_C$) and the right panel shows 
the case for $N_t=24$ ($T\approx 1.5T_C$).
For ease of viewing, the three data points in the right panel are slightly offset on the x-axis so that they do not overlap.
There is no particular finite-volume dependence on either side in all four channels. 
Although calculations based solely on different volumes do not seem to lead to a definitive conclusion, our results may suggest that the two-point functions $C(t)$ at both high and low temperatures are dominated by a single hadron state whose finite-volume spectral weight should be scaled as $1/V_s$. In addition, it is also found that the constant contribution is also scaled by $1/V_s$ in all four channels.

Next let us take a closer look at the scaling of the finite-volume spectral weight of the glueball ground state with respect to the spatial volume $V_s$. Taking into account the ``wrap-around effect'' for the glueball ground state, the two-point function $C(t)$ is expressed by
\begin{align}
C(t)=C_0^\prime+\frac{W_G}{\sinh[M_GN_t/2]}\cosh[M_G (t-N_t/2)],
\end{align}
where $W_G$ denotes the finite-volume spectral weight of the glueball ground state, while $C_0^\prime$
corresponds to the constant contribution. Recall that the spectral weight $W_G$ and the constant $C_0^\prime$, can be determined by setting $W_G = A \sinh[M_G N_t/2] C(0)$ and $C_0^\prime = C_0 C(0)$, where $C_0$, $A$, and $M_G$ are three fit parameters defined in Eq.~(\ref{Eq:FitForm}) and the value of $C(0)$ is also known.

\begin{figure}[b]
\includegraphics[width=0.4\linewidth,bb=0 0 792 612,clip]{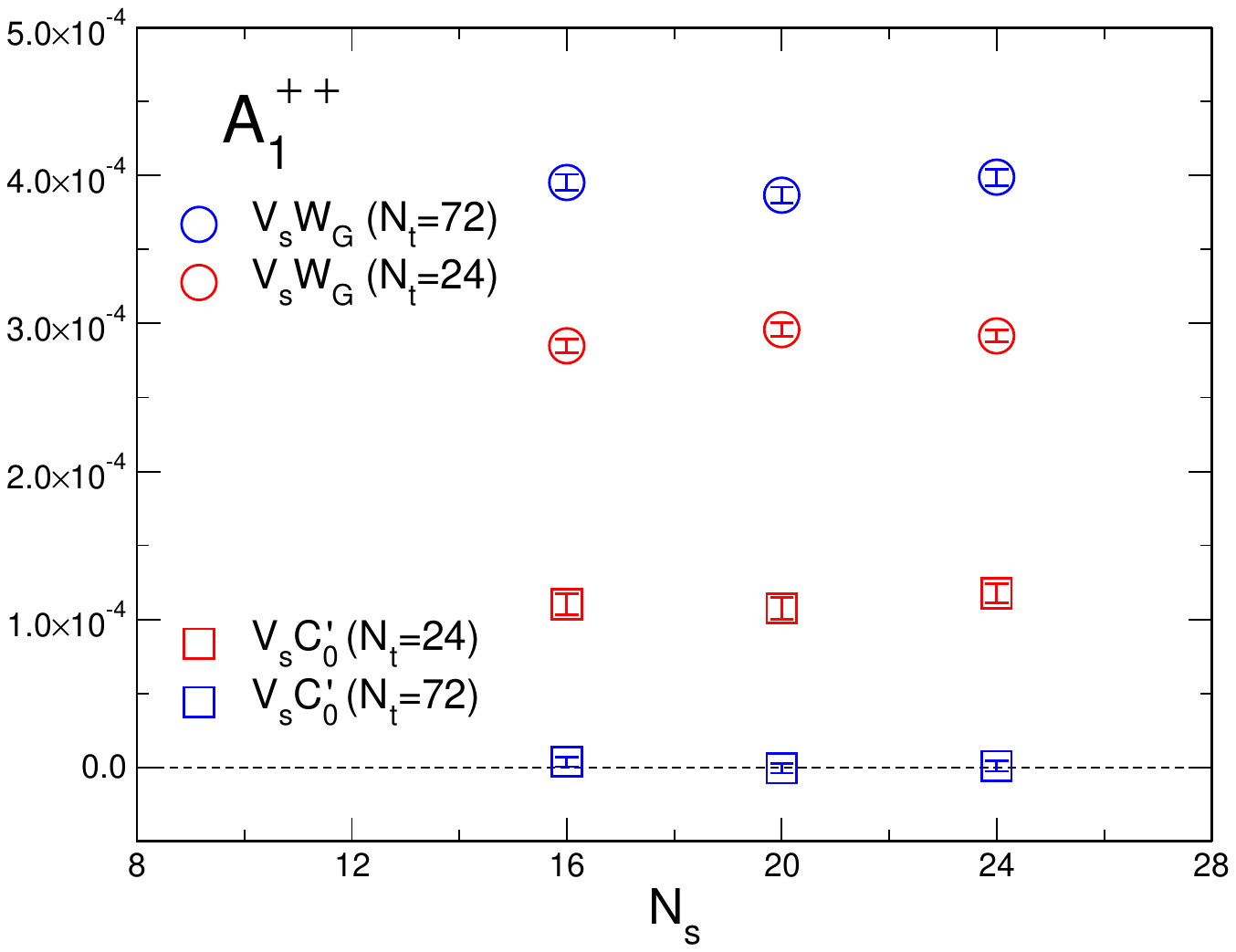}
\includegraphics[width=0.4\linewidth,bb=0 0 792 612,clip]{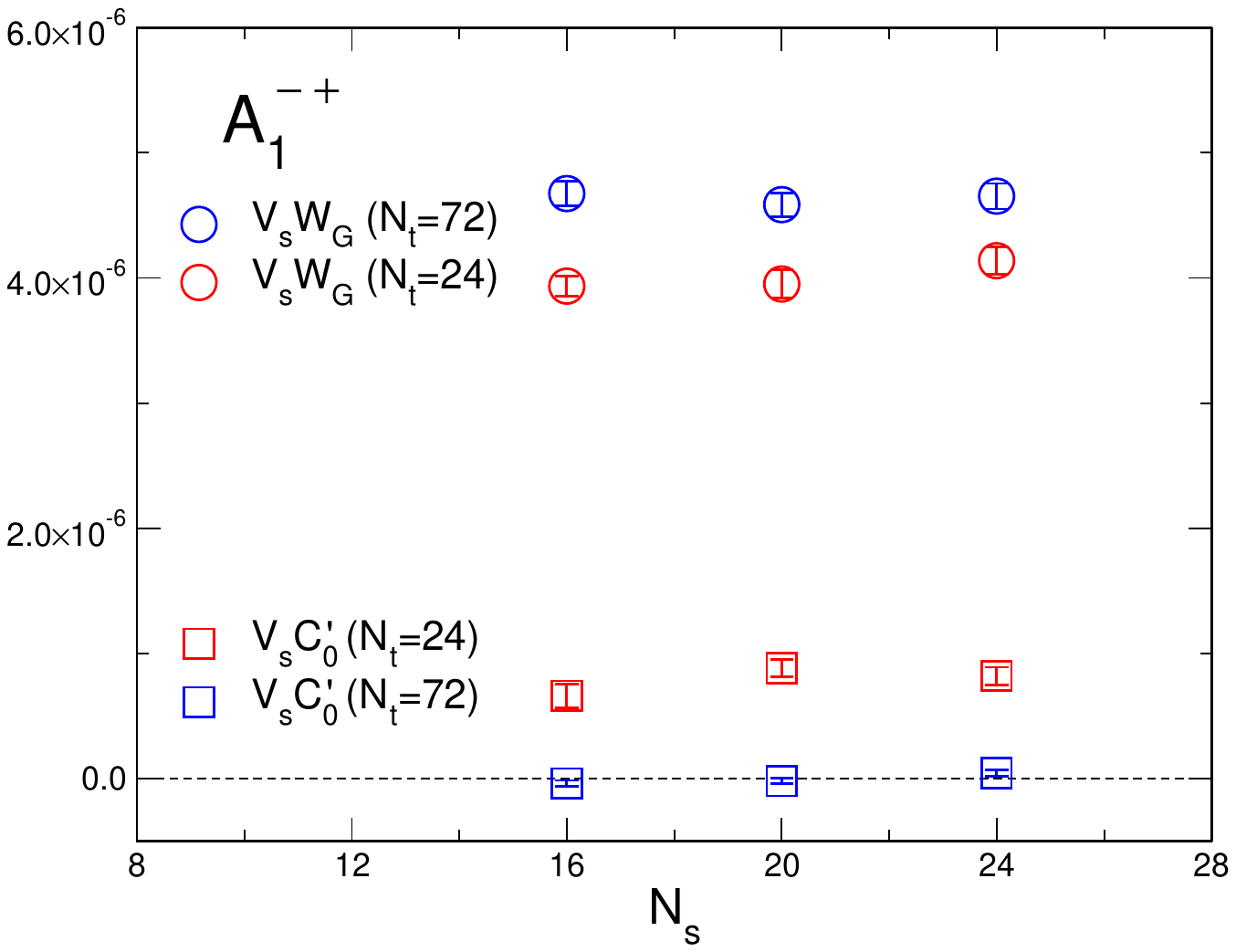}
\includegraphics[width=0.4\linewidth,bb=0 0 792 612,clip]{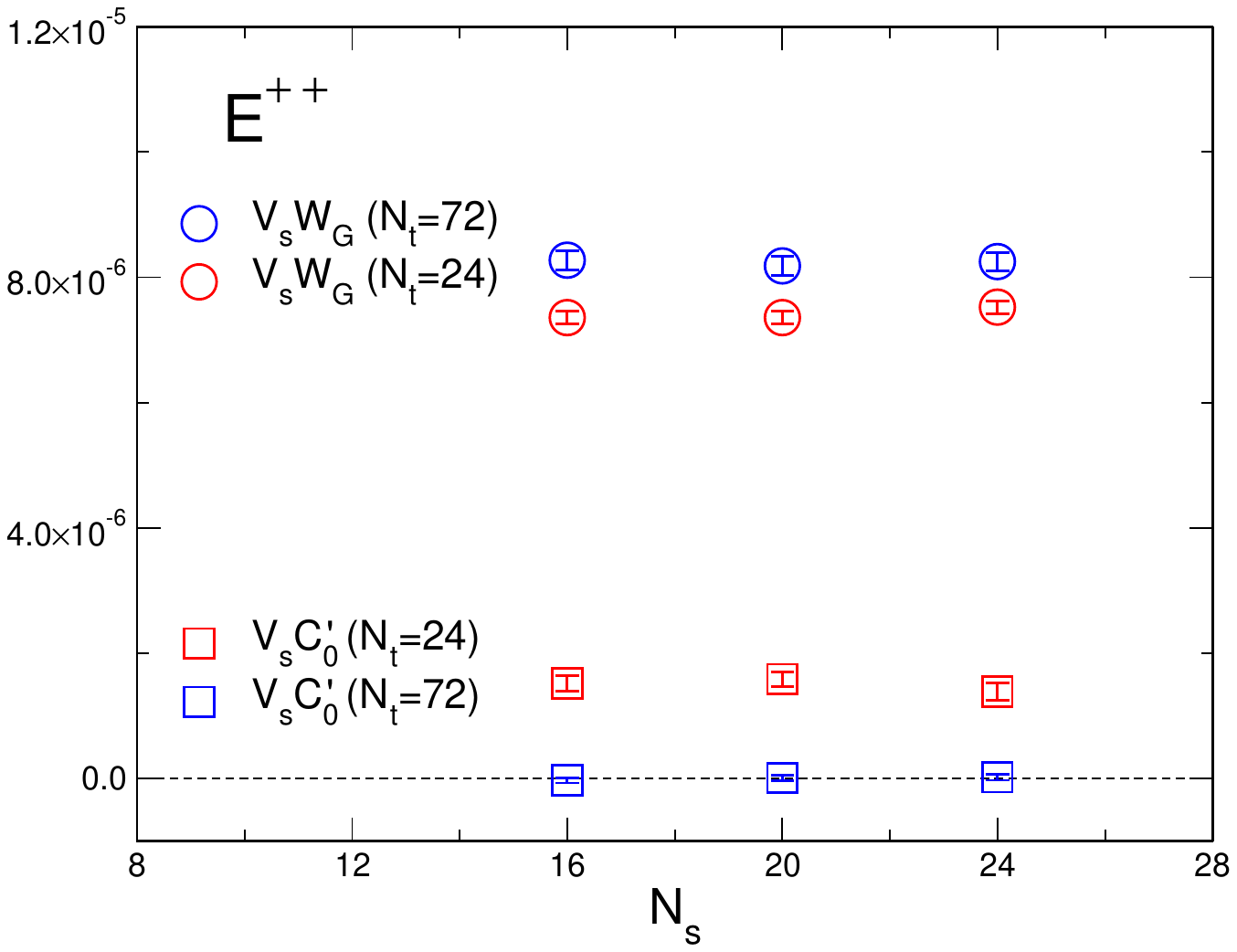}
\includegraphics[width=0.4\linewidth,bb=0 0 792 612,clip]{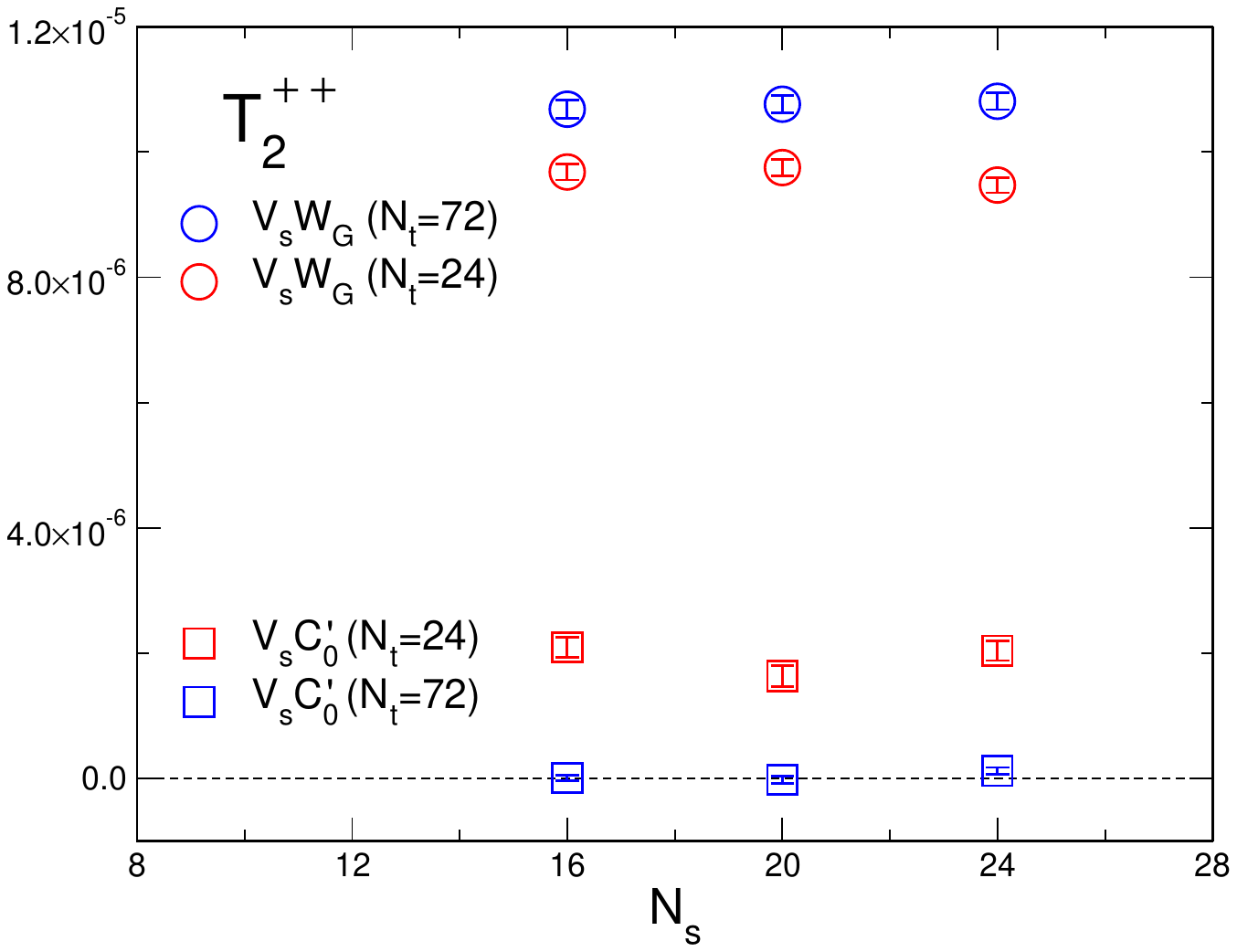}
 \caption{
Finite volume dependence of both the spectral weight of the glueball bound state ($W_G$)
and the constant term ($C_0^\prime$) on the low and high temperature sides. 
In each panel, $W_G$ (circle symbols) and $C_0^\prime$ (square symbols) scaled by the spatial volume $V_s=N_s^3$ are plotted as a function of $N_s$. 
The blue symbols denote the case for $N_t=72$ ($T\approx 0.5T_C$), while the red symbols denote the case for $N_t=24$ ($T\approx 1.5T_C$).
\label{fig:FV_SW}
}
\end{figure}

In Fig.~\ref{fig:FV_SW}, the finite-volume spectral weight $W_G$ and the constant contribution $C_0^\prime$
scaled by the spatial volume factor $V_s$ are plotted as a function of $N_s$. 
First of all, there is no significant dependence on $N_s$ for the glueball ground state 
(denoted by circle symbols) at either low or high temperatures in all four channels. 
This implies that the volume dependence assuming a single-hadron state persists even after the phase transition. Furthermore, the squared symbols demonstrate analogous volume scaling in all four channels for the constant contribution, which manifests exclusively after the phase transition.

%
%
%

%
%
%

%
%
%

%
%
\begin{table*}[h]
\centering
\caption{
Masses of the ground state of the $A_1^{++}$ and $A_1^{-+}$ irreps at finite temperature on $20^3 \times N_t$ lattices.\label{tab:fit_mass_T1}
}
\begin{ruledtabular} 
\begin{tabular}{c l c c c c c}
\hline
Irreps & $N_t$ & $C_0$ & $A$  & $a_t M_G$ & Fit range & $\chi^2/{\rm dof}$ \\
\hline
$A_1^{++}$ 
&72 &	$-0.0015(78)$& 0.0076(11) & 0.1534(41)	& [4,25] &1.06 \cr
&48 &	0.0189(105)	& 0.0392(44) & 0.1613(46)	& [4,18] &0.70 \cr
&40 & 	0.0251(163)	& 0.0799(101) & 0.1573(60)	& [4,18] &1.54 \cr
&36 &   0.0933(97)    & 0.0785(56) & 0.1725(36)       & [3,16] &1.67 \cr
&34 & 	0.1186(120)	&0.0863(73)  & 0.1753(46)  	& [3,16] &0.76 \cr
&32 & 	0.1490(114)	&0.0783(69)  & 0.1.901(52)	& [3,15] &1.29 \cr
&28 & 	0.1904(148)	&0.1025(105)  & 0.1934(66)	& [3,13] &1.21 \cr
&24 & 	0.2544(172)	&0.1096(133)  & 0.2127(91)	& [3,11] &0.74 \cr
&20 & 	0.3388(222)	&0.1120(190) & 0.2392(153)	& [3, 9] &1.06 \cr
\hline
$A_1^{-+}$ 
&72	& $-0.0038(45)$  &0.0003(1) &0.2425(71) &[4,25]	&0.77\cr
&48	& 0.0071(50)     &0.0042(5) &0.2539(55) &[3,20]		&1.04\cr
&40	& 0.0053(65)     &0.0126(15) &0.2498(62)&[3,20]	&1.53\cr
&36	&0.0112(53)      &0.0220(18) &0.2457(46)&[3,18]	&1.50\cr
&34	&0.0352(65)      &0.0239(22) &0.2550(56)&[3,17]	&0.98\cr
&32	&0.0497(80)      &0.0368(36) &0.2415(62)&[3,16]&1.11\cr
&28	&0.0976(96)      &0.0483(52) &0.2520(79)&[3,14]	&0.71\cr
&24	&0.1683(127)    &	0.0488(88) &0.2859(167)&[4,12]&1.45\cr
&20	&0.2053(162)    &0.0749(123) &0.2939(167)&[3, 9]&0.24\cr
\hline
\end{tabular}
\end{ruledtabular}
\end{table*}

%
%
\begin{table*}[h]
\centering
\caption{
Masses of the ground state of the $E^{++}$ and $T_2^{++}$ irreps at finite temperature on $20^3 \times N_t$ lattices.\label{tab:fit_mass_T2}
}
\begin{ruledtabular} 
\begin{tabular}{c l c c c c c}
\hline
Irreps & $N_t$ & $C_0$ & $A$  & $a_t M_G$ & Fit range & $\chi^2/{\rm dof}$ \\
\hline
$E^{++}$ 
&72&	0.0009(50)	&0.0005(1)	&0.2317(52)	 &[4,25]& 0.96\cr
&48&	0.0089(55)	&0.0065(7)	&0.2372(50)	 &[3,20]& 0.65\cr 
&40&	0.0041(77)	&0.0196(22)	&0.2290(58)	 &[3,18]& 0.72\cr	
&36&	0.0246(60)	&0.0254(24)	&0.2359(56)	 &[4,16]& 0.32\cr	
&34&	0.0314(77)	&0.0372(33)	&0.2298(53)	 &[3,16]& 1.54\cr
&32&	0.0530(88)	&0.0451(34)	&0.2306(57)	 &[3,14]& 1.66\cr
&28&	0.1105(104)	&0.0571(59)	&0.2415(73)	 &[3,14]& 0.73\cr
&24&	0.1697(125)	&0.0659(80)	&0.2644(101) &[3,12]& 1.50\cr
&20&	0.2000(180)	&0.0931(80)	&0.2771(150) &[3,10]& 0.65\cr
\hline
$T_2^{++}$ 
&72&	$-$0.0017(48)	&0.0005(1)	&0.2311(52)	&[3,28]&	1.65\cr
&48&	0.0041(59)	&0.0087(10)	&0.2239(48)	&[3,20]&	0.69\cr
&40&	$-$0.0017(82)	&0.0182(26)	&0.2335(76)	&[4,20]&	1.42\cr
&36&	0.0100(61)	&0.0316(24)	&0.2276(42)	&[3,18]&	1.06\cr
&34&	0.0403(73)	&0.0319(28)	&0.2392(53)	&[3,17]&	0.57\cr
&32&	0.0564(86)	&0.0452(33)	&0.2299(58)	&[3,15]&	1.63\cr
&28&	0.0834(110)	&0.0667(66)	&0.2330(69)	&[3,14]&	0.28\cr
&24&	0.1363(139)	&0.0785(95)	&0.2525(99)	&[3,12]&	1.34\cr
&20&	0.2285(166)	&0.0812(127)&0.2877(156)&[3,10]& 1.62\cr
\hline
\end{tabular}
\end{ruledtabular}
\end{table*}

\section{Summary}
\label{sec:SUMMARY}
We have studied the glueball properties at finite temperature from the temporal correlation in pure $SU(3)$ Yang-Mills theory using anisotropic lattice QCD. The deconfinement phase transition becomes first order for the pure $SU(3)$ Yang-Mills theory without quarks.
Accordingly, the glueball state, which is the sole hadronic excitation in this theory, permits direct study of how the mass modifications of hadrons stemming from the deconfinement phase transition. 

The previous studies~\cite{{Ishii:2001zq},{Ishii:2002ww},{Meng:2009hh}} have indeed shown that the glueball ground-states survive as a hadronic state after the deconfinement phase transition, and their masses decrease with temperature above the critical temperature $T_C$. However, the presence of the constant contribution appearing in the glueball two-point functions with respect to the deconfinement phase transition was not taken into account in their analysis. It has been observed that constant contributions have been evident only for mesonic correlators in the scalar and axis-vector channels~\cite{{Umeda:2007hy},{Ohno:2011zc}}.
However, given the underlying mechanism that governs
these contributions, it is reasonable to deduce that they are inevitable in the $C=+1$ glueball two-point functions. 

To confirm the existence of the constant contribution, 
we have compared the effective mass plots given by
two methods. One is the effective mass obtained from the original two-point function $C(t)$, and the other is the effective mass obtained by the time-differential one as $D(t)=\frac{1}{2}\left(C(t+1)-C(t-1)\right)$, where the constant contribution can be eliminated. It was observed that the effective masses of the two methods coincide just before the phase transition point, but a difference appears after the phase transition and then increases with increasing temperature $T$.

By performing the standard pole-mass analysis, which
includes the constant contribution not taken into account in the previous studies, it is concluded that the glueball ground-state masses in the three lowest-lying channels increase with temperature $T$ after the deconfinement phase transition. Furthermore, their normalized amplitudes, which correspond to
the strength of the spectral weights, are reduced with increasing $T$ after the deconfinement phase transition. Our result indicates that the true temperature dependence of the glueball mass above $T_C$ is opposite to the results of the previous studies~\cite{{Ishii:2001zq},{Ishii:2002ww},{Meng:2009hh}}.

We have also investigated whether the transition from a single-hadron state to the multi-gluon state occurred before and after the deconfinement phase transition, using simulations in three different volumes, and found that the volume dependence assuming a single-hadron state persists after the phase transition.
This seems to be consistent with the conclusion of the previous studies that the ``glueball-like modes'' survive up to around $T\approx 1.5 T_C$. A more definitive conclusion might be to check for spatial localization by, for example, examining wave functions of ``glueball-like modes''~\cite{{deForcrand:1991kc},{Liang:2014jta}}. 
We plan to conduct such a study in the future.


\begin{acknowledgments}

K. S. was supported by Graduate Program on Physics for the Universe (GP-PU)
of Tohoku University. 
Numerical calculations in this work were partially performed using Yukawa-21 
at the Yukawa Institute Computer Facility. 
This work was also supported in part by Grants-in-Aid for Scientific Research form the Ministry 
of Education, Culture, Sports, Science and Technology (No. 22K03612).

\end{acknowledgments}


\end{document}